\documentclass[showpacs,aps,floatfix,pra,reprint,superscriptaddress,nofootinbib]{revtex4-1}
\usepackage{graphicx}
\usepackage{amsmath}
\usepackage{amssymb} 
\usepackage{bm}
\usepackage{mathtools}
\usepackage{hyperref} \hypersetup{colorlinks=true,citecolor=blue,linkcolor=blue,urlcolor=blue}
\usepackage{listings}
\usepackage{multirow}
\usepackage[dvipsnames]{xcolor}
\usepackage{soul}
\usepackage{makecell} 
\usepackage{tabularx}
\usepackage{tabu} 
\usepackage{longtable} 
\usepackage{supertabular} 
\usepackage{enumitem}
\usepackage{xcolor} 
\usepackage{newverbs} 
\newverbcommand{\cverb}{}{} 

\DeclareFixedFont{\ttb}{T1}{txtt}{bx}{n}{8} 
\DeclareFixedFont{\ttm}{T1}{txtt}{m}{n}{8}  

\definecolor{pranab_green}{rgb}{0.31,0.53,0.10}
\definecolor{tmrwBlue}{rgb}{0.259,0.443,0.68.2}
\definecolor{tmrwRed}{rgb}{0.784,0.157,0.161}
\definecolor{tmrwGreen}{rgb}{0.443,0.549,0}
\definecolor{tmrwPurple}{rgb}{0.537,0.349,0.659}
\definecolor{tmrwAqua}{rgb}{0.243,0.6,0.624}
\definecolor{tmrwYellow}{rgb}{0.918,0.718,0}
\definecolor{tmrwOrange}{rgb}{0.871,0.576,0.373}
\definecolor{tmrwComment}{rgb}{0.557,0.565,0.549}

\newcommand\pythonstyle{\lstset{
language=Python,
basicstyle=\ttm,
otherkeywords={__init__, self},             
keywordstyle=\ttb\color{tmrwBlue},
emph={and,break,class,continue,def,yield,del,elif ,else,%
except,exec,finally,for,from,global,if,import,as,%
lambda,not,or,pass,print,raise,return,try,while,assert,with},
emphstyle=\ttb\color{tmrwPurple},    
emph={[2]},
emphstyle=[2]\ttb\color{tmrwBlue},
emph={[3] in },
emphstyle=[3]\ttb\color{tmrwAqua},
emph={[4]object,type,list,set,len,dict,tuple,str,repr,int,float},
emphstyle=[4]\ttb\color{tmrwYellow},
emph={[5]aflow_sym, pprint, Symmetry, json, subprocess, os, aflow_command, get_symmetry, get_edata, get_sgdata, loads, path, realpath},
emphstyle=[5]\ttm\color{tmrwBlue},
emph={[6]True, False, None},
emphstyle=[6]\ttm\color{tmrwOrange},
emph={[7]self},
emphstyle=[7]\ttm\color{tmrwRed},
stringstyle=\color{tmrwGreen},
morecomment=[s]{"""}{"""},
commentstyle=\color{tmrwComment}\ttm,
literate=
 {-}{{{-}}}1
 {<}{{{<}}}1,
frame=tb,                         
breaklines=true,
postbreak=\mbox{\textcolor{tmrwRed}{$\hookrightarrow$}\space},
showstringspaces=false            %
}}

\lstnewenvironment{python}[1][]
{
  \pythonstyle
  \lstset{#1}
}
{}

\newcommand\pythoninline[1]{{\pythonstyle\lstinline!#1!}}

\lstdefinelanguage{mylang}{
  basicstyle=\ttfamily,
  alsoletter=0123456789,
  alsodigit={.-}
}
\setlist[itemize]{noitemsep, topsep=0pt}
\newlist{myitemize}{itemize}{3}
\setlist[myitemize,1]{label=\textbullet,leftmargin=1em}
\setlist[myitemize,2]{label=--,leftmargin=1em}
\setlist[myitemize,3]{label=$\diamond$,leftmargin=1em}
\setlist[myitemize]{noitemsep, topsep=0pt}

\newcommand{\degrees}[0]{^\circ} 
\bibpunct{[}{]}{,}{n}{}{} 

\setstcolor{red} 

\def\AFLOW{{\small AFLOW}}
\def\AFLOWVERSION{{{{3.1.169}}}} 
\def\AFLOWICSDDATE{6 October 2017}
\def\QUANTUMESPRESSO{\textsc{Quantum {\small ESPRESSO}}}
\def\FHIAIMS{{\small FHI-AIMS}}
\def\ABINIT{{\small ABINIT}}
\def\AFLUX{{\small AFLUX}}
\def\RESTAPI{{\small REST-API}}
\def\API{{\small API}}

\def\VASP{{\small VASP}}
\def\INCAR{{\small INCAR}}
\def\OUTCAR{{\small OUTCAR}}
\def\ICSD{{\small ICSD}}
\def\CSD{{\small CSD}}
\def\OQMD{{\small OQMD}}

\def\CIF{\texttt{\small CIF}}
\def\SPF{\texttt{\small SPF}}
\def\POSCAR{\texttt{\small POSCAR}}

\def\AFLOWSYM{{\small AFLOW-SYM}}
\def\FINDSYM{{\small FINDSYM}}
\def\ADDSYM{{\small ADDSYM}}

\def\ITC{{\small ITC}}
\def\PLATON{{Platon}}
\def\SPGLIB{{Spglib}}
\def\AFLOWpi{{\small AFLOW$\pi$}}
\def\NOMAD{{NoMaD}}
\def\AFLOWCHECK{{\textcolor{pranab_green}{\checkmark}}}

\def\citeAFLOW{\cite{aflowPAPER,curtarolo:art110,curtarolo:art85,curtarolo:art63,curtarolo:art58,curtarolo:art57,curtarolo:art53,curtarolo:art49,monsterPGM,aflowANRL,aflowPI}}

\makeatletter \renewcommand\frontmatter@abstractwidth{\dimexpr\textwidth\relax} \makeatother  

\begin{document}

\title{\LARGE AFLOW-SYM: Platform for the complete, automatic and self-consistent symmetry analysis of crystals} 

\author{David Hicks}
\affiliation{Department of Mechanical Engineering and Materials Science, Duke University, Durham, North Carolina 27708, USA}
\affiliation{Center for Materials Genomics, Duke University, Durham, North Carolina 27708, USA}
\author{Corey Oses}
\affiliation{Department of Mechanical Engineering and Materials Science, Duke University, Durham, North Carolina 27708, USA}
\affiliation{Center for Materials Genomics, Duke University, Durham, North Carolina 27708, USA}
\author{Eric Gossett}
\affiliation{Department of Mechanical Engineering and Materials Science, Duke University, Durham, North Carolina 27708, USA}
\affiliation{Center for Materials Genomics, Duke University, Durham, North Carolina 27708, USA}
\author{Geena Gomez}
\affiliation{Department of Mechanical Engineering and Materials Science, Duke University, Durham, North Carolina 27708, USA}
\affiliation{Center for Materials Genomics, Duke University, Durham, North Carolina 27708, USA}
\author{Richard H. Taylor}
\affiliation{Department of Mechanical Engineering and Materials Science, Duke University, Durham, North Carolina 27708, USA}
\affiliation{Center for Materials Genomics, Duke University, Durham, North Carolina 27708, USA}
\affiliation{Department of Materials Science and Engineering, Massachusetts Institute of Technology, Cambridge, Massachusetts 02139, USA}
\author{\\Cormac Toher}
\affiliation{Department of Mechanical Engineering and Materials Science, Duke University, Durham, North Carolina 27708, USA}
\affiliation{Center for Materials Genomics, Duke University, Durham, North Carolina 27708, USA}
\author{Michael J. Mehl}
\affiliation{United States Naval Academy, Annapolis, Maryland 21402, USA}
\author{Ohad Levy}
\affiliation{Department of Mechanical Engineering and Materials Science, Duke University, Durham, North Carolina 27708, USA}
\affiliation{Center for Materials Genomics, Duke University, Durham, North Carolina 27708, USA}
\affiliation{Department of Physics, NRCN, P.O. Box 9001, Beer-Sheva 84190, Israel}
\author{Stefano Curtarolo}
\email[]{stefano@duke.edu}
\affiliation{Department of Mechanical Engineering and Materials Science, Duke University, Durham, North Carolina 27708, USA}
\affiliation{Center for Materials Genomics, Duke University, Durham, North Carolina 27708, USA}
\affiliation{Fritz-Haber-Institut der Max-Planck-Gesellschaft, 14195 Berlin-Dahlem, Germany}





\date{\today}

%
%
\begin{abstract}
  \noindent Determination of the symmetry profile of structures is a persistent challenge in materials science.
  Results often vary amongst standard packages, hindering autonomous
  materials development by requiring continuous user attention and
  educated guesses. 
  Here, we present a robust procedure for evaluating the complete suite
  of symmetry properties,
  featuring various representations for the point-, factor-, space groups, site symmetries, and Wyckoff positions.
  The protocol determines a system-specific mapping tolerance
  that yields symmetry operations entirely commensurate with fundamental crystallographic principles.
  The self consistent tolerance characterizes the effective spatial resolution of the reported atomic positions.
  The approach is compared with the most used programs and is successfully validated
  against the space group information provided for over 54,000 entries in the
  Inorganic Crystal Structure Database.
  Subsequently, a complete symmetry analysis is applied to all 1.7+ million entries of
  the \AFLOW\ data repository.
  The \AFLOWSYM\ package has been implemented in,
  and made available for, public use through the automated,
  \textit{ab-initio} framework \AFLOW.
\end{abstract}
\maketitle

%
%






\section{Introduction}

Symmetry fundamentally characterizes all crystals, establishing
a tractable connection between observed phenomena and the underlying
physical/chemical interactions.
Beyond crystal periodicity,
symmetry within the unit cell guides materials
classification~\cite{aflowANRL},
optimizes materials properties calculations,
and instructs structure enumeration methods~\cite{Buerger_JCP_1947,enum1}.
Careful exploitation of crystal symmetry has made possible the characterization
of electronic~\cite{curtarolo:art58}, mechanical~\cite{Toher_PRB_AGL_2014,curtarolo:art115},
and thermal properties~\cite{curtarolo:art114,curtarolo:art119,curtarolo:art125} in high-throughput fashion~\cite{nmatHT} --- giving
rise to large materials properties databases such as
\AFLOW~\citeAFLOW, \NOMAD~\cite{nomad}, 
Materials Project~\cite{materialsproject.org},
and \OQMD~\cite{oqmd.org}.
As these databases incorporate more properties and grow increasingly integrated,
access to rapid and consistent symmetry characterizations becomes of paramount importance.

Central to each symmetry analysis is the identification of spatial and angular tolerances, quantifying
the threshold at which two points or angles are considered equivalent.
These tolerances must account for numerical instabilities, and, more importantly, 
for atypical data stemming from finite temperature measurements or deviations in experimentally
measured values~\cite{lepage_jacryst_1987}.
Existing symmetry platforms --- such as \FINDSYM~\cite{stokes_findsym,Stokes_FROZSL_Ferroelectrics_1995}, 
\PLATON~\cite{platon_2003}, and \SPGLIB~\cite{spglib} 
{{--- all cater to different symmetry objectives, and thus address tolerance issues in unique ways.}}
\FINDSYM\ {{--- designed for ease-of-use ---}} acknowledges that its algorithms cannot handle noisy data, and it applies no 
treatments for ill-conditioned data~\cite{stokes_findsym}.  
The \PLATON\ geometry package, containing the subroutine \ADDSYM, allows a small percentage of candidate
atomic mappings to fail {{and attempts to capture missing higher symmetry descriptions~\cite{platon_2003}.
Lower symmetry descriptions in atomic coordinates can originate from 
{\bf i.} extraction issues with X-ray diffraction data (\textit{e.g.}, incorrectly identified crystal system, 
altered Laue class within a crystal system, and neglected inversion) and 
{\bf ii.} \textit{ab-initio} relaxations (\textit{e.g.}, lost internal 
translations)~\cite{Baur_actacrystB_1986,Herbstein_actacrystB_1982,Marsh_actacrystB_1983}.}} 
The \SPGLIB\ package applies independent tolerance scans within subroutines --- \textit{e.g.}, 
in its methods for finding the primitive cell (\verb|get_primitive|) and 
Wyckoff positions (\verb|ssm_get_exact_positions|) --- if certain crystallographic conventions are violated,
potentially yielding globally inconsistent symmetry descriptions~\cite{spglib}.
These packages present suggested default tolerance values that are largely arbitrary, and likely
justified \textit{a posteriori} on a limited test set.
In the general case, or in the event where these global defaults fail, 
the packages fall back on user-defined tolerances.
Unfortunately, it is difficult to compare results across packages outside of these default values because
tolerances are defined differently.
\FINDSYM\ and \SPGLIB\ both offer a tunable atomic mapping tolerance,
along with a lattice tolerance (\FINDSYM) and an angular tolerance (\SPGLIB);
whereas \PLATON\ has four separate input tolerances, each specific to a particular operation type.
Ultimately, these inconsistencies are symptomatic of an underlying inability
to appropriately address tolerances in symmetry analyses.

\begin{figure*}
  \begin{center}
    \includegraphics[width=0.995\textwidth]{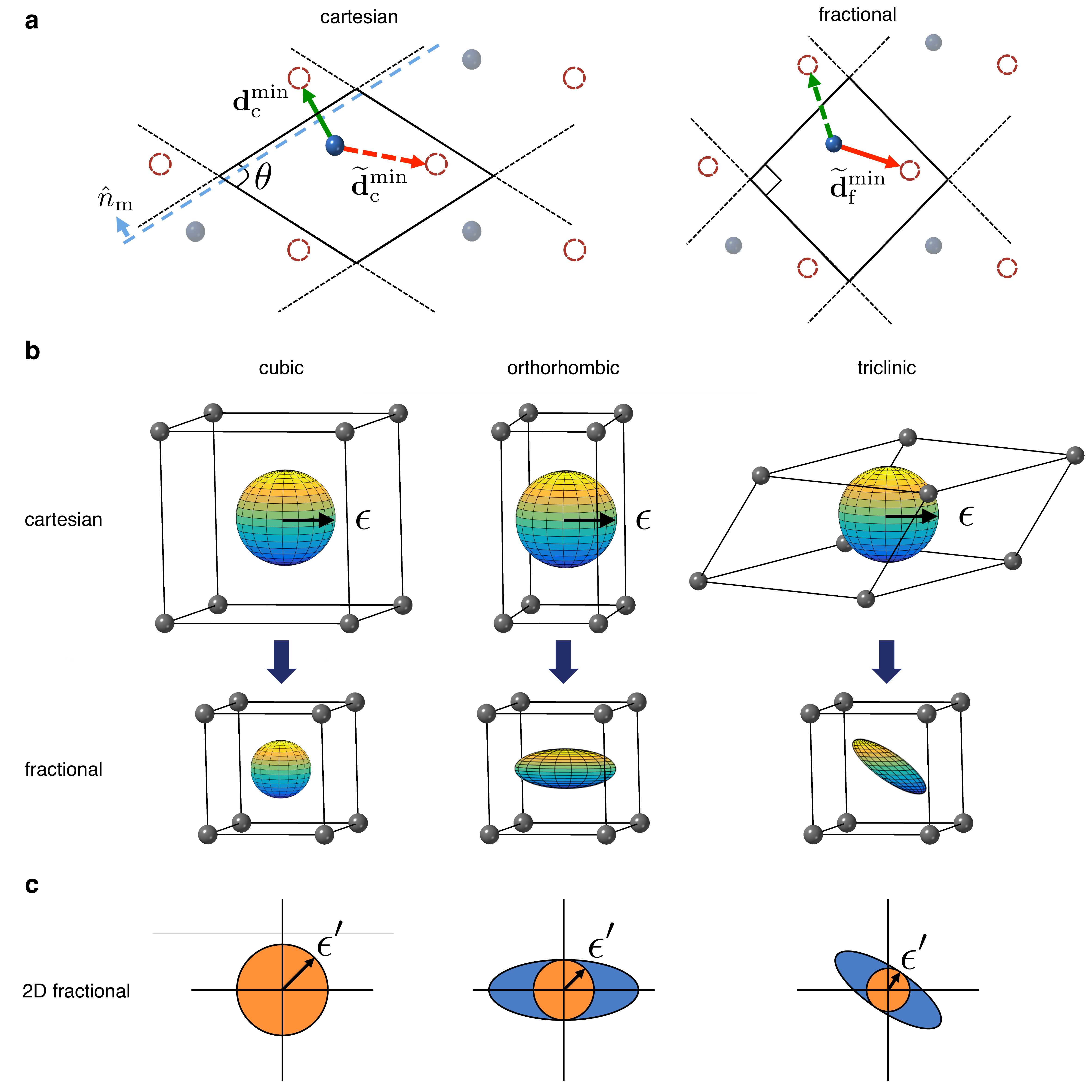}
    \caption{\small
      \textbf{Visualizations of space warping with a basis transformation.}
      (\textbf{a}) To validate a candidate mirror operation (described by $\hat{n}_{\mathrm{m}}$) on a crystal
      (blue atoms), the operation is applied to yield a transformed crystal (hollow orange atoms superimposed 
      on the original crystal).
      The true distance between the blue and orange atoms is resolved in cartesian space, indicated by
      the green $\mathbf{d}^{\mathrm{min}}_{\mathrm{c}}$ vector.
      However, the bring-in-cell method determines another periodic image to be closer, highlighted by the
      dashed red vector. 
      The mismatch is obscured in fractional space, where the red vector appears smaller than the green,
      indicated by $\mathbf{\widetilde{d}}^{\mathrm{min}}_{\mathrm{c}}$.
      (\textbf{b}) An atom is placed in the middle of the lattice with a surrounding sphere of radius $\epsilon$.
      Mapping occurs when the position of an atom transformed by a symmetry operator is within the sphere.
      The size and shape of the sphere is warped with a basis transformation (cartesian to fractional):
      uniform compression occurs in cubic cells,
      oblate compression in orthorhombic cells,
      non-uniform (sheared) compression in triclinic lattices.
      (\textbf{c}) 2D illustration of how the tolerance ($\epsilon$) warps in fractional space for cubic, 
      orthorhombic and triclinic lattices.  
      The orange circle with radius $\epsilon^{\prime}$ in fractional coordinates indicates the bounds of the safe mapping region, 
      independent of direction.
    }
    \label{fig:skew_examples_final}
  \end{center}
\end{figure*}

Managing input/output formats for these packages also presents a challenge.
\FINDSYM\ and \PLATON\ both read \CIF\ and \SPF\ files, which are particularly useful for structures deriving from
larger crystal structure databases, such as the 
{\underline{I}}norganic 
{\underline{C}}rystal 
{\underline{S}}tructure
{\underline{D}}atabase (\ICSD)~\cite{ICSD,ICSD3} and 
{\underline{C}}ambridge
{\underline{S}}tructural
{\underline{D}}atabase (\CSD)~\cite{Groom_CSD_2016}.
\PLATON\ also supports a few other useful input formats, while
\SPGLIB\ created its own input format.
Package-specific formats are useful for the developers, but create an 
unnecessary hurdle for the user to implement structure-file converters.
This is particularly problematic when package developers change the formats
of these inputs with new version updates, forcing the user to continuously adapt workflows/frameworks.
Additionally, all output formats are package-specific, with a medley of symmetry
descriptions and representations provided among the three.
The assortment of outputs presents yet another hurdle for users trying to 
build custom solutions for framework integration.
Furthermore, it forces users to become locked-in to these packages. 

These issues require extensive maintenance on the side of the user,
with little guarantee of the validity of the resulting symmetry descriptions.
In the case of large materials properties databases, 
providing such individual attention to each compound's symmetry description
becomes entirely impractical.
Herein, we present a robust symmetry package implemented in the automated, \textit{ab-initio} 
computational framework \AFLOW, known as \AFLOWSYM.
The module delivers a complete symmetry analysis
of the crystal, including the symmetry operations for the lattice point group, reciprocal
space lattice point group, factor group, crystal point group, dual of the crystal point group, symmetrically equivalent atoms,
site symmetry, and space group (see Appendix~\ref{sec:symmetry_groups} for an overview of symmetry groups).
Moreover, it provides general crystallographic descriptions
including the space group number and label(s), Pearson symbol, Bravais lattice type and variations, Wyckoff positions,
and standard representations of the crystal.  
The routine employs an adaptive, structure-specific tolerance scheme
capable of handling even the most skewed unit cells.
By default, two independent symmetry procedures are applied, 
enabling corroboration of the characterization.
The scheme has been tested on 54,000 \ICSD\ compounds in the {\sf aflow.org} repository, 
showing substantial improvement in characterizing space groups and lattice types compared to 
other packages.
Along with a standardized text output, \AFLOWSYM\ presents
the results in \underline{J}ava\underline{S}cript \underline{O}bject \underline{N}otation (JSON)
for easy integration into different workflows.  
The software is completely written in C++ and it can be compiled in
{\small UNIX}, Linux, and MacOSX environments using the gcc/g++ suite of compilers.
The package is open-source and is available under the {\small GNU-GPL} license.
An \AFLOWSYM\ Python module is also available to facilitate integration with other workflow packages, {\it e.g.},
\AFLOWpi~\cite{aflowPI,curtarolo:art93} and \NOMAD~\cite{nomad}. 
Thus, \AFLOWSYM\ serves as a robust one-stop symmetry shop for the
materials science community.

\begin{figure}
  \begin{center}
    \includegraphics[width=0.995\linewidth]{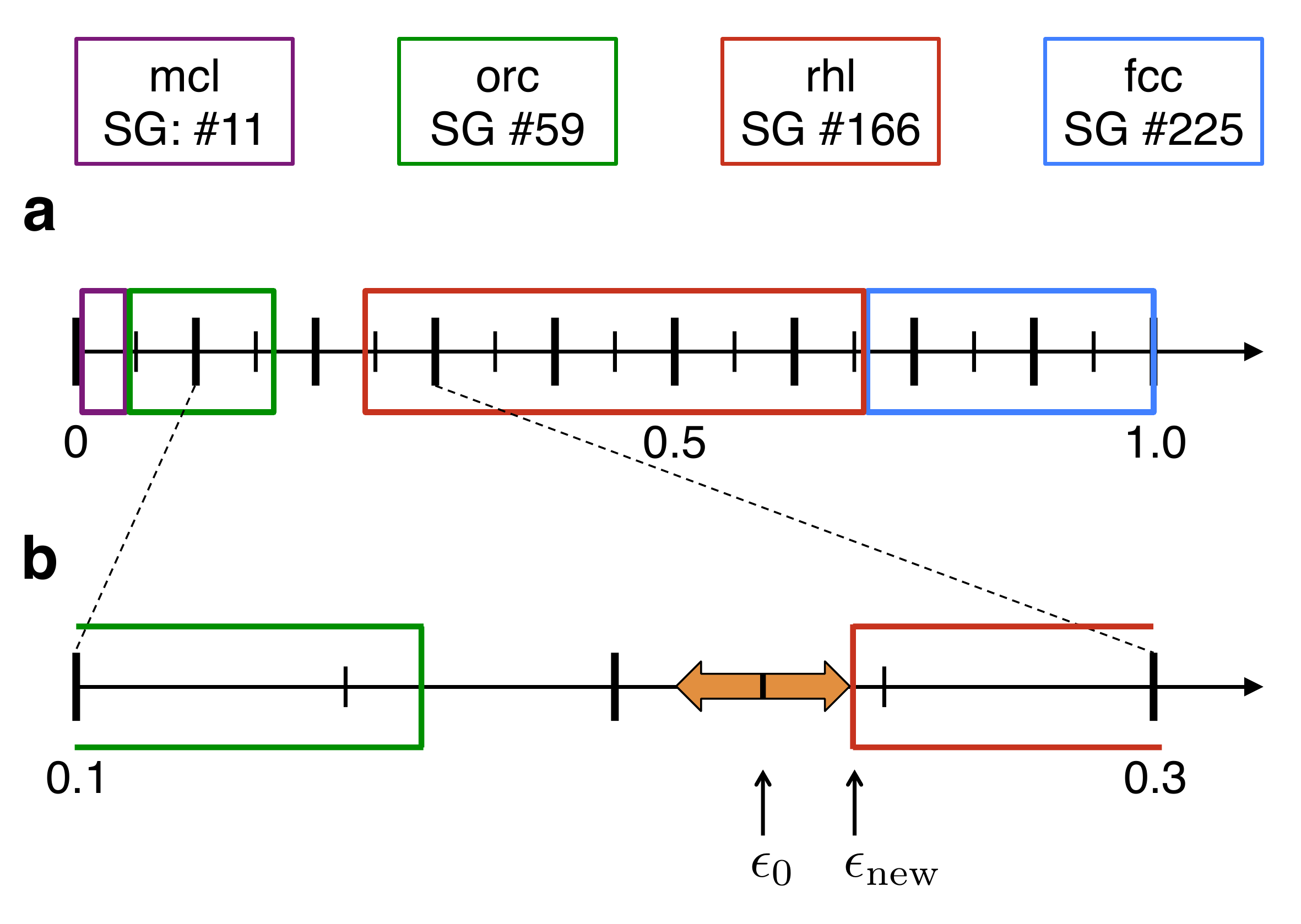}
    \caption{
      \textbf{Variation of space group with mapping tolerance
        for AgBr (\ICSD\ \#56551) 
        as determined by \AFLOWSYM.}
      (\textbf{a}) Space groups and tolerance ranges identified are as follows (ascending order):
      $1.0\times10^{-6} - 4.0855\times10^{-2}~\mathrm{\AA}$ is monoclinic (SG \#11),
      $4.08556\times10^{-2} - 1.64186\times10^{-1}~\mathrm{\AA}$ is orthorhombic (SG \#59),
      $2.46281\times10^{-1} - 6.69605\times10^{-1}~\mathrm{\AA}$ is rhombohedral (SG \#166),
      and $6.69606\times10^{-1} - 1.0~\mathrm{\AA}$ is face-centered cubic (SG \#225).
      (\textbf{b}) A gap is highlighted between $1.64186\times10^{-1} - 2.46281\times10^{-1}~\mathrm{\AA}$
      where no consistent space group is identified.
      The orange arrows illustrate how the algorithm scans possible tolerances to find
      the closest consistent space group.
    }
    \label{fig:tolerance_spectrum_scan}
  \end{center}
\end{figure}

\section{Methods}

\subsection{Periodic boundary conditions in skewed cells}
\label{subsec:PBC}
Analyzing the symmetry of materials involves determining the full set of their isometries. 
Algorithmically, candidate symmetry operators are applied to a set of atoms and validated if 
{\bf i.} distances between atoms and their transformed counterparts 
are within a mapping tolerance $\epsilon$, and 
{\bf ii.} the mappings are isomorphic (one-to-one).
For convenience, $\epsilon$ is defined in units of a Euclidean space
--- Angstroms in this case.
An explicit mapping function is defined, indicating whether atom mappings are successful:
\begin{equation}
  \mathrm{map}_{\mathrm{atom}}\left(\mathbf{d}_{\mathrm{c}}\right) = 
  \begin{cases}
    \mathrm{true} & \text{if $\left\Vert\mathbf{d}_{\mathrm{c}}\right\Vert< \epsilon$} \\
    \mathrm{false} & \text{otherwise}
  \end{cases},
  \label{eq:mapping_function}
\end{equation}
where $\mathbf{d}_{\mathrm{c}}$ is the cartesian distance vector between an atom and a transformed atom. 
Symmetries of the crystal are discovered when successful isomorphic mappings --- given by Equation~(\ref{eq:mapping_function}) --- 
exist between all of the original and transformed atoms.
Under periodic boundary conditions, the 
minimum distance for the mapping function is identified by considering 
equivalent atoms of nearby cells (so-called method of images~\cite{hloucha_minimumimage_1998}):
\begin{equation}
  \mathbf{d}^{\mathrm{min}}_{\mathrm{c}} = \min_{n_a,n_b,n_c}\left(\mathbf{d}_{\mathrm{c}}+n_a\mathbf{a}+n_b\mathbf{b}+n_c\mathbf{c}\right), \\
\end{equation}
where $\mathbf{a}$, $\mathbf{b}$, $\mathbf{c}$ are the lattice vectors;
$n_a$, $n_b$, $n_c$ are the indices of neighboring cells;
and $\mathbf{d}^{\mathrm{min}}_{\mathrm{c}}$ is the globally optimal cartesian distance vector.
In the simplest case of a purely orthorhombic cell,
the approach requires exploration of the 
26 surrounding unit cells $\left(-1\leq n_a,n_b,n_c \leq 1\right)$.
However, additional neighboring cells should be considered with increased skewness of the lattice vectors
(see Section~\ref{subsec:radiusspherelattice}),
making it prohibitively expensive.
Instead, many algorithms minimize the distance vector through a greedy, bring-in-cell approach~\cite{hloucha_minimumimage_1998}.
Working with fractional coordinates, each component $i$ of the distance vector $\mathbf{d}_{\mathrm{f}}$
is minimized using the nearest-integer function ($\mathrm{nint}$~\cite{hloucha_minimumimage_1998}):
\begin{equation}
  \begin{gathered}
    \label{eq:nint}
    \mathbf{\widetilde{d}}_{\mathrm{f},i}^{\mathrm{min}} = \mathbf{d}_{\mathrm{f},i}-\mathrm{nint}\left(\mathbf{d}_{\mathrm{f},i}\right) \quad \forall i, \\ 
    \mathbf{\widetilde{d}}_{\mathrm{c}}^{\mathrm{min}} = \mathbf{L}\mathbf{\widetilde{d}}_{\mathrm{f}}^{\mathrm{min}},
  \end{gathered}
\end{equation}
where $\mathbf{L}$ is the column-space matrix representation of the lattice, and  
$\mathbf{\widetilde{d}}^{\mathrm{min}}_{\mathrm{c}}$ is $\mathbf{\widetilde{d}}^{\mathrm{min}}_{\mathrm{f}}$ converted
to cartesian coordinates for the mapping determination in Equation~(\ref{eq:mapping_function}).

While a convenient shortcut, the bring-in-cell minimum distance is not generally equivalent to the 
globally optimized distance:
$\mathbf{\widetilde{d}}^{\mathrm{min}}_{\mathrm{c}} \neq \mathbf{d}^{\mathrm{min}}_{\mathrm{c}}$
(see Figure~\ref{fig:skew_examples_final}(a)).
A component-by-component minimization of the distance vector assumes independent basis vectors (no skewness)
and neglects potentially closer images only considered by exploring all neighboring cells.
{{Occurrence}} of a distance mismatch depends on the lattice type 
and compromises the integrity of the mapping determination.
The issue becomes particularly elusive in fractional coordinates, 
where the size and shape of the cell is warped to yield 
a unit cube as shown in Figure~\ref{fig:skew_examples_final}(b).
The greater the anisotropy of the lattice, the larger the warping.
Figure~\ref{fig:skew_examples_final}(c) illustrates how the tolerance changes between cartesian and fractional space.
In the general case, a spherical tolerance in cartesian coordinates warps into an ellipsoid in fractional coordinates.
Hence, the criteria for successful mappings in fractional space are direction dependent unless the distance is sufficiently small, {\it i.e.},
within the circumscribed sphere of radius $\epsilon^{\prime}$ (highlighted in orange).
Distances within $\epsilon^{\prime}$ in fractional space always map within $\epsilon$ in cartesian space, but a
robust check (global optimization) is needed for larger distances to account for the extremes of the ellipsoid.
Since most distances outside of $\epsilon^{\prime}$ do not yield mappings, such a robust check is generally wasteful.
Instead, more useful insight can be garnered from Figure~\ref{fig:skew_examples_final}(c):
tolerances sufficiently bounded by the skewness can still yield a proper mapping determination using
the inexpensive bring-in-cell algorithm.

The goal is to define an upper tolerance threshold to safely ensure that the bring-in-cell minimum
($\mathbf{{\widetilde{d}}}^{\mathrm{min}}_{\mathrm{c}}$) and global minimum ($\mathbf{{d}}^{\mathrm{min}}_{\mathrm{c}}$)
yield the same mapping results, in spite of a distance mismatch:
\begin{equation}
  \begin{gathered}
    \left. \mathrm{map}_{\mathrm{atom}}\left(\mathbf{d}^{\mathrm{min}}_{\mathrm{c}}\right) \equiv
      \mathrm{map}_{\mathrm{atom}}\left(\mathbf{\widetilde{d}}^{\mathrm{min}}_{\mathrm{c}}\right)\,\middle|\,
      \epsilon < \left\Vert\mathbf{d}^{\mathrm{min}}_{\mathrm{c}}\right\Vert 
    \right. , \\
    \left. \forall \, \mathbf{d}_{\mathrm{c}} \,\middle|\, 
      \mathbf{d}^{\mathrm{min}}_{\mathrm{c}} \neq \mathbf{\widetilde{d}}^{\mathrm{min}}_{\mathrm{c}}\right. .
  \end{gathered}
\end{equation}
A mismatch is encountered when the image identified by the bring-in-cell method is not the optimal neighbor, therefore:
$\mathbf{{d}}^{\mathrm{min}}_{\mathrm{c}} \le \mathbf{{\widetilde{d}}}^{\mathrm{min}}_{\mathrm{c}}$.
A suitable threshold needs to overcome the difference between the two methods for all
mismatch possibilities, {\it i.e.}, $\epsilon$ would need to be below $\mathbf{{d}}^{\mathrm{min}}_{\mathrm{c}}$ or above
$\mathbf{{\widetilde{d}}}^{\mathrm{min}}_{\mathrm{c}}$ to yield a consistent mapping determination.
A threshold greater than $\mathbf{{\widetilde{d}}}^{\mathrm{min}}_{\mathrm{c}}$ 
is ruled out to ensure that $\epsilon$ is always smaller than the minimum interatomic distance $\left(d^{\mathrm{nn(min)}}_{\mathrm{c}}\right)$,
making it possible to distinguish nearest-neighbors.
To find a tolerance in the remaining region $\left(\epsilon < ||\mathbf{{d}}^{\mathrm{min}}_{\mathrm{c}}||\right)$,
the largest mismatch possible should be addressed directly, which yields the smallest $\mathbf{{d}}^{\mathrm{min}}_{\mathrm{c}}$
and thus the most restrictive bound on the tolerance.
Given the angles between the lattice vectors $\left(\alpha,\beta,\gamma\right)$,
a maximum skewness is defined as
\begin{equation}
  \xi_{\mathrm{max}} = \max \left(\cos{\alpha},\cos{\beta},\cos{\gamma}\right),
  \label{eqn:max_skew}
\end{equation}
where the cosine of the angle derives from the normalized, off-diagonal
terms of the metric tensor.
$\xi_{\mathrm{max}}$ ranges from $[0,1)$, where 
$\xi_{\mathrm{max}}=0$ characterizes a perfectly orthorhombic cell.
A suitable maximum mapping tolerance is heuristically defined as
\begin{equation}
  \epsilon_{\mathrm{max}} = \left(1-\xi_{\mathrm{max}}\right) d^{\mathrm{nn(min)}}_{\mathrm{c}},
  \label{eqn:max_tol}
\end{equation}
which appropriately reduces $d^{\mathrm{nn(min)}}_{\mathrm{c}}$ --- an absolute upper bound for the 
tolerance to maintain resolution between atoms --- with increasing skewness.
The form of the coefficient ($1-\cos \theta $) decays quickly with basis vector overlap 
(on the order of $\theta^2$), ensuring a safe enveloping bound.
Tolerances well below $\epsilon_{\mathrm{max}}$ should yield the correct mapping determination with
the bring-in-cell approach (in spite of a distance mismatch);
otherwise, the global minimization algorithm should be employed:
\begin{equation}
  \label{eqn:map_choice}
  \mathbf{d}^{\mathrm{map}}_{\mathrm{c}} = 
  \begin{cases}
    \mathbf{L}\mathbf{\widetilde{d}}_{\mathrm{f}}^{\mathrm{min}} & \text{if $\epsilon \ll \epsilon_{\mathrm{max}}$} \\
    \displaystyle\min_{n_a,n_b,n_c}\left(\mathbf{d}_{\mathrm{c}}+n_a\mathbf{L}_{a}+n_b\mathbf{L}_{b}+n_c\mathbf{L}_{c}\right) & \text{otherwise} \\
  \end{cases}.
\end{equation}
To demonstrate the robustness of $\epsilon_{\mathrm{max}}$, 
extreme hypothetical cases are presented in Appendix \ref{sec:metric_warping}. 

\begin{figure*}
  \begin{center}
    \includegraphics[width=0.995\textwidth]{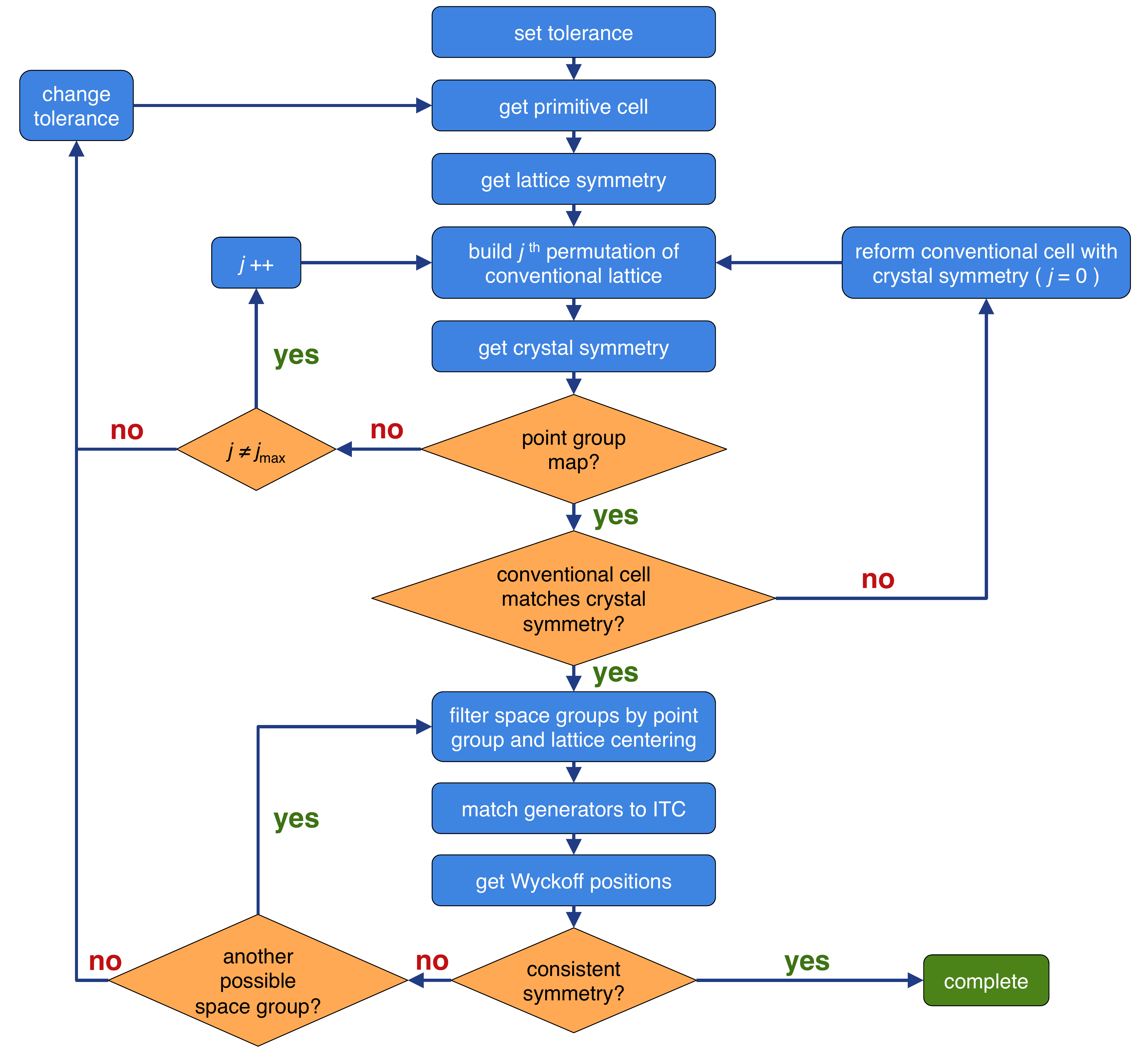}
    \caption{\textbf{Workflow for the algorithm converting a structure to the standard representation
        as defined by the International Tables for Crystallography.}
      Functions are represented by blue rectangles, and validation schemes
      by orange diamonds.
    }
    \label{fig:aflowSG_flowchart}
  \end{center}
\end{figure*}

\begin{figure*}
  \begin{center}
    \includegraphics[width=0.995\textwidth]{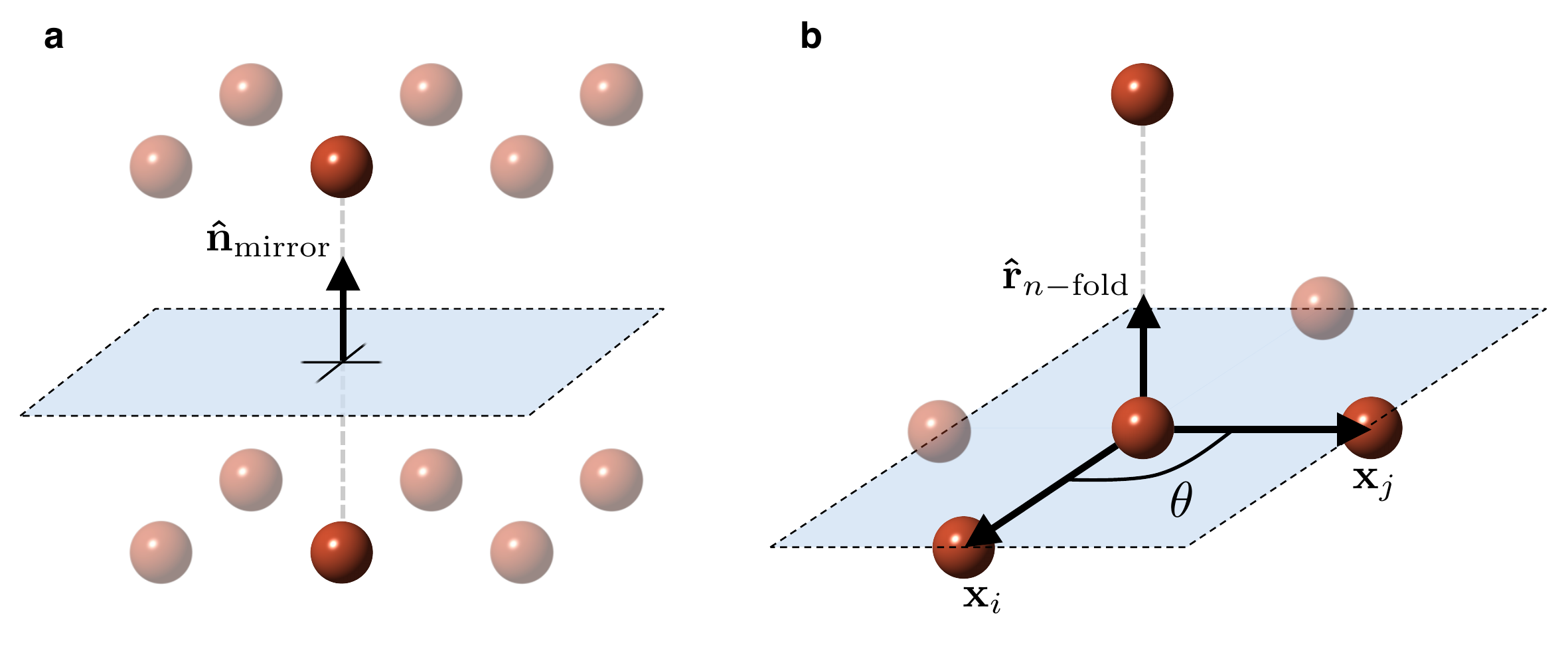}
    \caption{\textbf{Minimum symmetry identifiers of the lattice system:}
      (\textbf{a}) mirror operations, 
      (\textbf{b}) $n$-fold rotations.
      The resulting lattice vectors are denoted by gray dotted lines.
    }
    \label{fig:lattice_ops}
  \end{center}
\end{figure*}

\subsection{Adaptive tolerance scheme} 
While $\epsilon_{\mathrm{max}}$ offers a practical upper tolerance bound for the choice of the
distance minimization algorithm, it offers no insight for choosing a specific tolerance.
Of course, there are fundamental constraints, such as the minimum interatomic distance 
and the precision of the input structure parameters:
$\epsilon_{\mathrm{precision}}<\epsilon<d^{\mathrm{nn(min)}}_{\mathrm{c}}$,
but these can span over several orders of magnitude, throughout which a variety of results are possible.
Figure~\ref{fig:tolerance_spectrum_scan} illustrates the different space groups that may be assigned to  
AgBr (\ICSD\ \#56551)~\footnote{{\sf http://www.aflow.org/material.php?id=56551}} with various tolerance choices (the \ICSD\ reports space group \#11).
Interestingly, adjacent space group regions show non-isomorphic subgroup relations: 
between space groups \#59 and \#11 and between \#225 and \#166.
Of particular concern is the gap highlighted in 
Figure~\ref{fig:tolerance_spectrum_scan}(b) between space group regions \#166 and \#59.
Not surprisingly, these space groups share no subgroup relations.
These gaps represent problematic regions where noise in the structural information 
interferes with determination of satisfied symmetry operations, yielding profiles inconsistent
with any possible space group.
Rather than an \textit{a posteriori} selection of the symmetry elements to include in the analysis, we employ 
an adaptive tolerance approach.
A radial tolerance scan is performed surrounding the initial input tolerance $\epsilon_{0}$
to overcome the ``{\it confusion}'' region, as shown in Figure~\ref{fig:tolerance_spectrum_scan}(b).
With each adjustment of the tolerance, the algorithm updates and validates all symmetry properties and
operations, yielding a globally consistent profile and an effective spatial resolution for the structure.

To fully characterize a structure's symmetry, \AFLOWSYM\ employs two major symmetry procedures.
The first calculates the symmetry of the crystal in the
\underline{I}nternational \underline{T}ables
for \underline{C}rystallography \cite{tables_crystallography,tables_crystallography_A1,bilbao,bilbao2} (\ITC) conventional cell, 
yielding the space group and Wyckoff positions. 
The second resolves the symmetry profile of the structure in the original (input) representation, including:
the lattice point group, reciprocal lattice point group, factor group, crystal point group, dual of the crystal point group, space group,
inequivalent and equivalent atoms, and the site symmetry. 
While both routines can be employed independently, the two are combined in \AFLOWSYM\ by default, affording
additional validation schemes to ensure a stricter consistency.

Ultimately, the combination of the tolerance scan and integrated workflow (with robust validation schemes)
ensures the automatic determination of a consistent symmetry profile.
While the option remains for a user defined tolerance (with and without the scan),
\AFLOWSYM\ heuristically defines two default tolerances values:
\verb|tight| ($\epsilon_{\mathrm{tight}} = d^{\mathrm{nn(min)}}_{\mathrm{c}}/100$) and 
\verb|loose| ($\epsilon_{\mathrm{loose}} = d^{\mathrm{nn(min)}}_{\mathrm{c}}/10$).
Generally, an expected symmetry profile (perhaps from experiments) can be found in
either of the two tolerances.
If no tolerance is defined, \AFLOWSYM\ defaults to the tight tolerance.
The tolerance chosen for the analysis is compared against $\epsilon_{\mathrm{max}}$ to identify 
the required minimization technique to yield consistent mappings (see Equation~(\ref{eqn:map_choice})).
Overall, the \AFLOWSYM\ tolerance scheme has been validated against 
54,000 \ICSD\ entries, and subsequently applied
to all 1.7 million entries stored in the {\sf aflow.org} repository.
The symmetry results can be retrieved from the \AFLOW\ repository via the \RESTAPI~\cite{curtarolo:art92} 
or the \AFLUX\ Search-\API~\cite{aflux}. 

\subsection{Tolerance types and conversions}
Outside of mapping distances, there are a number of relevant quantities for which an
equivalence criterion is required, \textit{e.g.}, lattice vectors, axes, angles, and symmetry operations.
Instead of defining separate tolerances for each, \AFLOWSYM\ leverages the single
spatial tolerance, converting quantities to cartesian distances whenever possible.
For vector quantities such as lattice vectors and axes, the difference is taken, converted to
the cartesian form (if necessary), and the Euclidean norm of the resulting vector is compared to the spatial tolerance.
For angles, each angle $\theta_i$ is converted to a straight-line distance $d_i$: 
\begin{equation}
  d_i=\bar{x}_i\sin\left(\theta_i\right),
\end{equation}
where $\bar{x}_i$ is the average length of the angle-defining vectors in cartesian space.
The two straight-line distances are subtracted and compared with the input spatial tolerance.
To compare rotation matrices for a particular lattice, each matrix is transformed into its fractional form, 
resulting in two integer matrices that can be matched exactly.

\subsection{International Tables for Crystallography standard representation} 
One strategy for uncovering a structure's symmetry profile is to convert it to a standard
form, such as the one defined by the \ITC.
In this representation, the symmetry operations, space group, and Wyckoff positions are 
well tabulated, mitigating the computational expense involved in combinatorial operation searches.
To efficiently explore the possibilities, the algorithm exploits the lattice symmetry
to resolve the crystal symmetry, from which the conventional cell is defined.
The full workflow is illustrated in Figure~\ref{fig:aflowSG_flowchart}.

First, the algorithm finds a primitive representation of the crystal (of which there are many) by
exploring possible internal translations forming a smaller lattice~\cite{tables_crystallography}.
To optimize the search, only the vectors between the least frequently occurring atomic species are considered.
The translation vector should preserve cell periodicity, and the resulting reduced representation should
conserve the stoichiometry.

Next, the symmetry of the lattice is determined by calculating the mirror and $n$-fold rotation operations.
The primitive cell is expanded from -1 to 1 in each direction~\cite{lepage_jacryst_1987}
and combinations of lattice points are considered for defining the following:
{\bf i.} mirror operations characterized by a plane $\left(\mathrm{normal~}\hat{\mathbf{n}}_{\mathrm{mirror}}\right)$
between two lattice points about which half the lattice points can be reflected onto the other, and
{\bf ii.} $n$-fold rotations $\left(n\in\{2, 3, 4, 6\}\right)$ described by an axis
$\left(\hat{\mathbf{r}}_{n\mathrm{-fold}}\right)$ and angle $\left(\theta\right)$ such that a rotation about 
$\hat{\mathbf{r}}_{n\mathrm{-fold}}$ by $\theta$ yields an isomorphic mapping of lattice points.
The two types of operations are illustrated in Figure~\ref{fig:lattice_ops}.

\renewcommand{\arraystretch}{2.0}
\begin{table*}
  \centering
  \begin{tabular}{||l||r|r|rl||}
    \hline
    \multirow{ 2}{*}{\shortstack[c]{lattice/crystal \\ system}} & \multicolumn{1}{c|}{\multirow{ 2}{*}{\# mirrors}} & 
                                                                                                                      \multicolumn{1}{c|}{\multirow{ 2}{*}{\# $n$-fold rotations}} & \multicolumn{2}{c||}{\multirow{ 2}{*}{conventional cell}} \\
                                                                & & & & \\
    \hline
    \hline
    cubic & 9 & 3 (four-fold)& $\mathbf{a}$,$\mathbf{b}$,$\mathbf{c}$ : & \makecell[l]{parallel to three equivalent four-fold axes} \\
    \hline
    hexagonal & 7 & 1 (three-/six-fold) & \makecell[r]{$\mathbf{c}$ : \\ $\mathbf{a}$,$\mathbf{b}$ : \\ \\ }  & \makecell[l]{parallel to three-/six-fold axis \\
    parallel to mirror axes \\
    ($|\mathbf{a}|\equiv|\mathbf{b}|$ and $\beta=120^{\circ}$)} \\
    \hline
    tetragonal & 5 & 1 (four-fold) & \makecell[r]{$\mathbf{c}$ : \\ $\mathbf{a}$,$\mathbf{b}$ : \\ \\ } & \makecell[l]{parallel to four-fold axis \\
    parallel to mirror axes \\
    ($|\mathbf{a}|\equiv|\mathbf{b}|$ and $\beta=90^{\circ}$)} \\
    \hline
    rhombohedral & 3 & 1 (three-/six-fold) & \makecell[r]{$\mathbf{c}$ : \\ $\mathbf{a}$,$\mathbf{b}$ : \\ \\ }  & \makecell[l]{parallel to three-/six-fold axis \\
    parallel to mirror axes \\
    ($|\mathbf{a}|\equiv|\mathbf{b}|$ and $\beta=120^{\circ}$)} \\
    \hline
    orthorhombic & 3 & - & $\mathbf{a}$,$\mathbf{b}$,$\mathbf{c}$ : & parallel to three mirror axes \\
    \hline
    monoclinic & 1 & - & \makecell[r]{$\mathbf{b}$ : \\ \\ $\mathbf{a}$,$\mathbf{c}$ : \\ \\ }  & \makecell[l]{parallel to mirror axis \\
    (unique axis) \\
    parallel to two (choice of three) \\ smallest translations
    perpendicular to $\mathbf{b}$~\cite{tables_crystallography}} \\
    \hline
    triclinic & 0 & - & $\mathbf{a}$,$\mathbf{b}$,$\mathbf{c}$ : & same as original lattice \\
    \hline
  \end{tabular}
  \caption{
    \textbf{Conventional cell construction rules based on symmetry operations.}
  }
  \label{tab:conventional_cell_rules}
\end{table*}
\renewcommand{\arraystretch}{1.0}

The cardinality of each operation type defines the lattice system, as detailed in Table~\ref{tab:conventional_cell_rules}.
If the lattice and crystal systems are the same, the characteristic vectors
of the lattice operators $\left(\hat{\mathbf{n}}_{\mathrm{mirror}}\mathrm{~and~}\hat{\mathbf{r}}_{n\mathrm{-fold}}\right)$
and corresponding lattice points
define the lattice vectors of the conventional cell (also outlined in Table~\ref{tab:conventional_cell_rules}).
For all cases, these lattice vectors are used to construct an initial conventional cell.
The aim is to find a conventional cell whose corresponding symmetry operators (tabulated in the \ITC\ in 
Table 11.2.2.1~\cite{tables_crystallography}) are validated for the crystal, which can have symmetry 
equal to or less than the lattice.
If a mismatch in cardinality is encountered, permutations of the lattice vectors are attempted.
Should a mismatch remain after all permutations have been exhausted, the conventional cell is 
reformed to reflect the crystal symmetry.
The reformed cell is chosen based on the observed cardinality of the symmetry operations (refer again to 
Table~\ref{tab:conventional_cell_rules}).

The resulting crystal point group set and internal translations (lattice centerings) are then used to filter
candidate space groups.
To pin down a space group exactly, the symmetry elements of the crystal are matched to the \ITC\ generators ---
the operations generating symmetrically equivalent atoms for the general Wyckoff position~\cite{tables_crystallography}.
However, a shift in the origin may differentiate the two sets of operators --- a degree of freedom
that should be addressed carefully.
The appropriate origin shift should transform the symmetry elements to the \ITC\ generators, 
thus forming a set of linear equations.
Consider two symmetrically equivalent atom positions ($\mathbf{x}$ and $\mathbf{x'}$) in the crystal,
\begin{equation}
  \mathbf{x'} = \mathbf{Ux} + \mathbf{t},
  \label{eqn:sym_equiv_points}
\end{equation}
where $\mathbf{U}$ and $\mathbf{t}$ are the fixed-point and translation operations, 
respectively, between the two atoms.
An origin shift $\mathbf{\mathcal{O}}$ relates these positions to those listed in the \ITC:
\begin{align}
  \label{eqn:origin_shift1a}
  \mathbf{x}_{\mathrm{ITC}} &= \mathbf{x} + \mathbf{\mathcal{O}}, \\
  \label{eqn:origin_shift1b}
  \mathbf{x'}_{\mathrm{ITC}} &= \mathbf{x'} + \mathbf{\mathcal{O}}.
\end{align}
Applying $\mathbf{U}$ to Equation~(\ref{eqn:origin_shift1a}) and subtracting it from
Equation~(\ref{eqn:origin_shift1b}) yields
\begin{equation}
  \mathbf{x'}_{\mathrm{ITC}} - \mathbf{U}\mathbf{x}_{\mathrm{ITC}} = \mathbf{x'} + \mathbf{\mathcal{O}} - \mathbf{Ux} - \mathbf{U\mathcal{O}}.
  \label{eqn:origin_shift2}
\end{equation}
The \ITC\ translation $\mathbf{t}_{\mathrm{ITC}}$ and the crystal translation $\mathbf{t}$ are related via
\begin{equation}
  \mathbf{t}_{\mathrm{ITC}} = \mathbf{t} + \mathbf{\mathcal{O}} - \mathbf{U\mathcal{O}}.
  \label{eqn:internal_shift}
\end{equation}
Combining Equations~(\ref{eqn:origin_shift2})-(\ref{eqn:internal_shift}) and incorporating 
Equations~(\ref{eqn:origin_shift1a})-(\ref{eqn:origin_shift1b})
produces the following system of equations:
\begin{equation}
  \left(\mathbf{I} - \mathbf{U}\right)\mathbf{\mathcal{O}} = \left(\mathbf{t}_{\mathrm{ITC}} - \mathbf{t}\right),
  \label{eqn:shift_system_of_equations}
\end{equation}
where $\mathbf{I}$ is the identity.
Equation~(\ref{eqn:shift_system_of_equations}) must be solved for each generator, often resulting in
an overdetermined system.
Periodic boundary conditions should also be considered when solving the system of equations, as solutions
may reside in neighboring cells.
If a commensurate origin shift is not found, the next candidate space group is tested.

With the shift into the \ITC\ reference frame, the Wyckoff positions are identified by grouping
atoms in the conventional cell into symmetrically equivalent sets.
These sets are compared with the \ITC\ standard to identify the corresponding Wyckoff coordinates,
site symmetry designation, and letter.
The procedure to find the origin shift is similarly applied to determine any Wyckoff parameters ($x$, $y$, $z$).
For some space groups, the Wyckoff positions only differ by an internal translation (identical site symmetries),
introducing ambiguity in their identification.
In these cases, \AFLOWSYM\ favors the Wyckoff scheme producing the smallest enumerated Wyckoff lettering.

After finding the Wyckoff positions, the algorithm is complete.  
\AFLOWSYM\ returns the space group, conventional cell, and Wyckoff positions in the \ITC\ standard representation.

\begin{figure*}
  \begin{center}
    \includegraphics[width=0.995\textwidth]{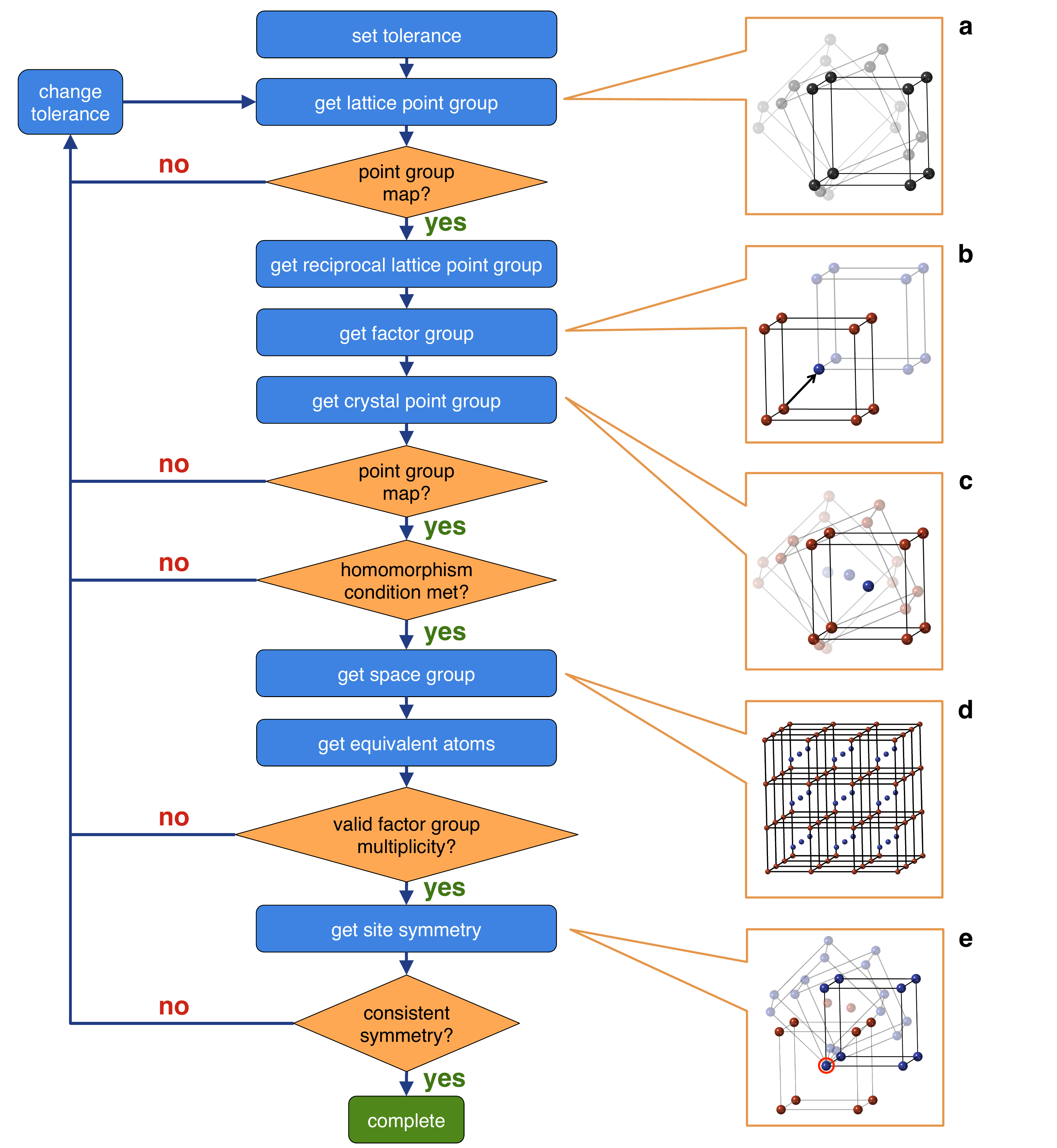}
    \caption{\textbf{Workflow for the algorithm calculating the symmetry operations of the system in its
        original representation.}
      Functions are represented by blue rectangles, and validation schemes
      by orange diamonds.
    }
    \label{fig:full_sym_flowchart}
  \end{center}
\end{figure*}

\subsection{Input orientation symmetry algorithm} 
The standard conventional cell representation described in the previous section affords easy access to the full symmetry profile
of the structure.
Nevertheless, other representations, such as the \AFLOW\ standard
primitive representation~\cite{curtarolo:art58},
are often preferred for reducing the computational cost of subsequent
calculations/analyses, 
such as density functional theory calculations \cite{curtarolo:art58}.
While conversions are always possible, as was done to find the standard conventional cell,
in practice it introduces errors in the structural parameters, becoming particularly problematic
in ``{\it confusion}'' tolerance regions (Figure~\ref{fig:tolerance_spectrum_scan}) and 
tolerance-sensitive algorithms, {\it e.g.}, calculations of force
constants \cite{curtarolo:art67,curtarolo:art125}.
To mitigate the need for error-accumulating conversions, 
a general-representation symmetry algorithm is also incorporated in \AFLOWSYM.
The integration of the two symmetry algorithms affords additional validation schemes
that combat ``{\it confusion}'' tolerance regions and ensures an overall stricter consistency.
The full workflow of this algorithm is outlined in Figure~\ref{fig:full_sym_flowchart}.
For descriptions of the different symmetry groups, refer to Appendix~\ref{sec:symmetry_groups}.

First, the point group of the lattice is calculated by finding all identical lattice cells
of an expanded grid (see Section~\ref{subsec:radiusspherelattice}).
The unique set of matrices that transform the rotated cells to the original cell
define the lattice point group, as depicted in Figure~\ref{fig:full_sym_flowchart}(a).
The search first considers all lattice points within a radius no smaller than that of a sphere
encapsulating the entire unit cell.
These points define the candidate lattice vectors (origin to lattice point), and those
not of length $a$, $b$, or $c$ (lattice vector lengths of original cell) are eliminated.
Next, all combinations of these candidate lattice vectors are considered, eliminating sets
by matching the full set of lattice parameters (lattice vector lengths and angles of the original cell).
The transformation matrix is calculated as
\begin{equation}
  \mathbf{U}_{\mathrm{c}}=\mathbf{L}\left(\mathbf{L}^{\prime}\right)^{-1}
\end{equation}
where $\mathbf{U}_{\mathrm{c}}$ is the cartesian form of the transformation (rotation) matrix, 
$\mathbf{L}$ is the original, column-space matrix representation of the lattice, and 
$\mathbf{L}^{\prime}$ is the rotated lattice.
The fractional form of the transformation matrix $\left(\mathbf{U}_{\mathrm{f}}\right)$ is similarly derived replacing
$\mathbf{L}$ and $\mathbf{L}^{\prime}$ with their fractional counterparts (the fractional form of $\mathbf{L}$
is trivially the identity matrix).

The calculation of the lattice point group allows rapid determination of its reciprocal space counterpart, 
describing the point group symmetry of the Brillouin zone.
The transformation of symmetry operators is straight-forward, following standard basis change rules in dual spaces.
A contragredient transformation converts the real-space form of the operator to its reciprocal counterpart,
which is trivial for the cartesian form of the operator (orthogonal matrix):
\begin{equation}
  \label{eqn:contragredient}
  \begin{aligned}
    \mathbf{V}_{\mathrm{c}} &= \left(\mathbf{U}_{\mathrm{c}}^{-1}\right)^{\mathrm{T}} = \mathbf{U}_{\mathrm{c}}, \\
    \mathbf{V}_{\mathrm{f}} &= \left(\mathbf{U}_{\mathrm{f}}^{-1}\right)^{\mathrm{T}}, \\
  \end{aligned}
\end{equation}
where $\mathbf{U}_{\mathrm{c}}$/$\mathbf{U}_{\mathrm{f}}$ and $\mathbf{V}_{\mathrm{c}}$/$\mathbf{V}_{\mathrm{f}}$ 
are the cartesian/fractional forms of the symmetry operator in real and reciprocal spaces, respectively~\cite{sands_vectorstensorscrystal_1982}.

Next, the coset representatives of the factor group are determined, 
characterizing the symmetry of the unit cell.
These operations are characterized by a fixed-point rotation (lattice point group) and an internal translation
that yield an isomorphic mapping among the atoms.
The smallest set of candidate translation vectors can be found among atoms of the least frequently occurring species.
This symmetry description is represented by Figure~\ref{fig:full_sym_flowchart}(b).

The point group of the crystal is then extracted from the coset representatives of the factor groups.
By exploiting the homomorphism --- or isomorphism for primitive cells --- between the factor group and the crystal point group,
the internal translations of the coset representatives are removed and the unique elements yield the crystal point group.
This is portrayed in Figure~\ref{fig:full_sym_flowchart}(c).
The dual space counterpart of the crystal point group is 
derived by performing the contragredient transformation, as shown in Equation~(\ref{eqn:contragredient}).

The space group operations are similarly derived from the coset representatives of the factor group.
The space group describes the symmetry of the infinitely periodic crystal, resulting from the
propagation of the unit cell symmetry throughout the lattice.
A finite set of space group operators are generated by applying the lattice translations to each of the
coset representative operations out to a specified radius.
The operation is depicted in Figure~\ref{fig:full_sym_flowchart}(d).

The coset representatives of the factor group also resolve the symmetrically equivalent (Wyckoff) atoms.
Atoms that are symmetrically equivalent map onto one another through a coset representative operation.
This organization is convenient for calculating the site symmetry of the crystal.  
The site symmetry, or site point group, are exposed by centering the reference frame 
onto each atomic site and applying the operations of the
crystal point group, as illustrated in Figure~\ref{fig:full_sym_flowchart}(e).
To expedite this process, the site symmetries are explicitly calculated for all inequivalent atoms.
They are then propagated to equivalent atoms with the appropriate change of basis (dictated by the coset representative
mapping the inequivalent atom to the equivalent atom).

\subsection{Consistency of symmetry} 
There are a finite number of operation sets that a crystal can exhibit~\cite{giacovazzofundamentals_1992}.
A set of symmetry operations outside of those allowed by crystallographic group theory
are attributed to noisy data, thus warranting the adaptive tolerance scan.
Numerous symmetry rules are validated throughout the \AFLOWSYM\ routines.  
The list of consistency checks are indicated below.
\begin{enumerate}
\item Point group (lattice/crystal) contains (at the very least) the identity element.
\item Point group (lattice/crystal) matches 1 of 32 point groups.
\item Coset representative of the factor group is an integer multiple of the crystallographic point group (homomorphic/isomorphic condition).
\item Space group symbol decomposes into crystallographic point group symbol by removing translational components
  (with the exception of derivative structures).
\item Number of symmetrically equivalent atoms is divisible by the ratio of the number of operations in the factor and crystal point groups.
\item Space group and Wyckoff positions match \ITC\ convention~\cite{tables_crystallography}.
\end{enumerate}

\subsection{Exploring the atomic environment} 
\label{subsec:radiusspherelattice}
A description of the local atomic environments in a crystal is 
required for determination of 
atom coordination and atom/lattice mappings.
Depending on the cell representation, an expansion is generally warranted for sufficient exploration of the nearest neighbors.
Here, an algorithm is outlined for determining the number of neighboring cells to explore in order to capture the 
local environment within a given exploration radius ($r_{\mathrm{sphere}}$).
In \AFLOWSYM, the default exploration radius is the largest distance between any two lattice points in a single unit cell.
First, the normal of each pair of lattice vectors is calculated and scaled to be of length $r_{\mathrm{sphere}}$,
\textit{e.g.},
$\mathbf{n}_1=r_{\mathrm{sphere}}\cdot{\mathbf{b}\times\mathbf{c}}/{\left\Vert\mathbf{b}\times\mathbf{c}\right\Vert}$,
where $\mathbf{b}$ and $\mathbf{c}$ are lattice vectors.
Next, the scaled normals are converted to the basis of the lattice, \textit{e.g.},
$\mathbf{n}_1^{\prime}=\mathbf{L}^{-1}\mathbf{n}_1$, where $\mathbf{L}$ is the column-space matrix representation of the lattice.
The magnitude (rounded up to the nearest integer) of the $i^{\mathrm{th}}$ component of the $\mathbf{n}_i^{\prime}$ vector 
reveals the pertinent grid dimensions $\left(d_1,d_2,d_3\right)$.
A uniform sphere of radius $r_{\mathrm{sphere}}$ centered at the origin fits
within a 3-D grid spanning $\left[-d_i,d_i\right]$.

\renewcommand{\arraystretch}{1.5}
\begin{table*}
  \centering
  \begin{tabular}{||l||r|r|r|r||}
    \hline
    \multicolumn{1}{||c||}{\multirow{ 2}{*}{package}} & \multirow{ 2}{*}{\shortstack[c]{\# space group \\ mismatches}} & \multirow{ 2}{*}{\shortstack[c]{\# lattice \\ mismatches}} 
                                                      & \multirow{ 2}{*}{\shortstack[c]{\# crystal system \\ mismatches}} & \multirow{ 2}{*}{\shortstack[c]{\# space group \\ not found}} \\
                                                      & & & & \\
    \hline
    \hline
    \AFLOWSYM\ & {{834}} & {{420}} & {{377}} & 0 \\
    \hline
    \SPGLIB\ & 10,022 (3,389)& 9,644 (2,917)& 9,523 (2,832)& 0 (0)\\
    \hline
    \FINDSYM\ & 3,540 (1,067)& 3,066 (531)& 2,982 (483)& 127 (156)\\
    \hline
    \PLATON & {{3,000 (1,217)}}& 1,092 (588)& 1,083 (486)& 195 {{(1,351)}}${^\ddagger}$\\ 
    \hline
  \end{tabular}
  \caption{
    \textbf{Mismatch counts between reported and calculated space groups for entries in the \ICSD.} 
    The test set is comprised of 54,015 \ICSD\ entries stored
    in the {\sf aflow.org} repository, as of \AFLOWICSDDATE.
    The columns indicate the number of entries whose space group, lattice type, and crystal family do not match
    those reported by the \ICSD.
    The results using the user-defined/non-default tolerance values
    for \SPGLIB, \FINDSYM, and \PLATON\ are shown in parentheses.  
    For more details, refer to the Supplementary Information.
    The superscript $^{\ddagger}$ indicates 2 entries for which the space group calculation exceeded 48 hours.
    }
  \label{table:symmetry_errors}
\end{table*}
\renewcommand{\arraystretch}{1.0}

\begin{figure*}
  \begin{center}
    \includegraphics[width=0.995\textwidth]{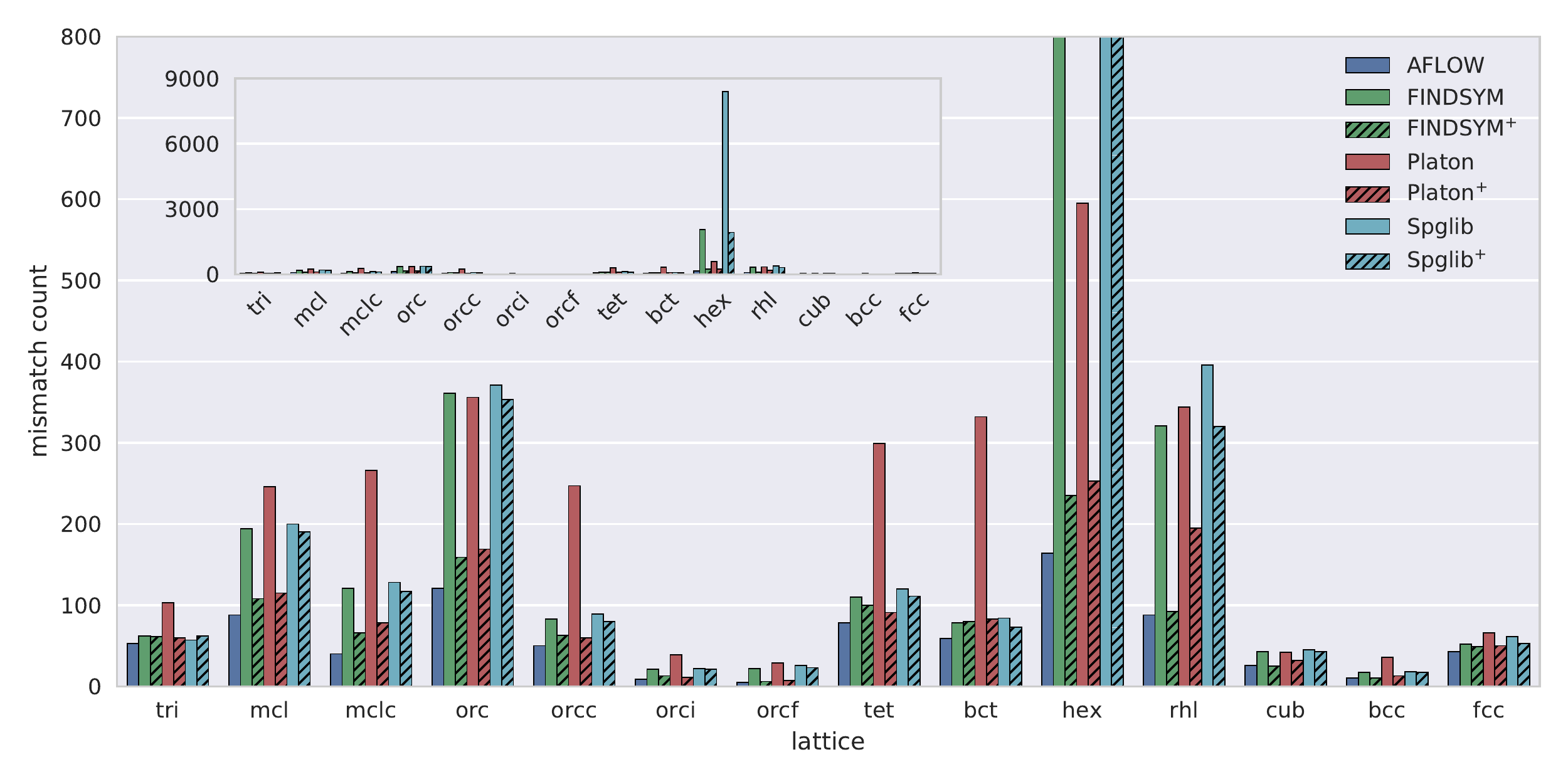}
    \caption{\textbf{Breakdown of space group mismatches with the \ICSD\ organized by lattice type.}
      The lattice types are derived from the space group number reported in the \ICSD.
      The superscript  $^{+}$ indicates the results using the user-defined/non-default tolerance values.
    }
    \label{fig:sg_inconsistencies}
  \end{center}
\end{figure*}

\section{Results}

Highlighted here are benchmarks to compare the various standard symmetry packages:
\AFLOWSYM, \SPGLIB, \FINDSYM, and \PLATON.
The results are calculated with the most recent versions available for download:
\begin{itemize}
\item \AFLOW\ version \AFLOWVERSION,
\item \SPGLIB\ version {{1.10.2.4}},
\item \FINDSYM\ version {{5.1.0}},
\item \PLATON\ version {{30118}}.
\end{itemize}
The default tolerances are employed as reported by the authors:
\begin{itemize}
\item \AFLOWSYM: $\epsilon_{\mathrm{tight}}=d^{\mathrm{nn(min)}}_{\mathrm{c}}/100$,
\item \SPGLIB: $\verb|symprec|=1\times10^{-5}$~\AA, \verb|angle_tolerance| derives from 
  \verb|symprec| --- default listed on web page~\cite{spglib_symmetry_html},
\item \FINDSYM: $\epsilon_{\mathrm{lattice}}=1\times10^{-5}$~\AA, $\epsilon_{\mathrm{atomic~position}}=1\times10^{-3}$~\AA --- default 
  from web interface~\cite{findsym_web_interface},
\item \PLATON: $\epsilon_{\mathrm{metric}}=1.00\degrees$, $\epsilon_{\mathrm{rotation}}=0.25$~\AA, 
  $\epsilon_{\mathrm{inversion}}=0.25$~\AA, $\epsilon_{\mathrm{translation}}=0.25$~\AA~\cite{lepage_jacryst_1987}.
\end{itemize}
Alternative tolerances values are also used for \SPGLIB, \FINDSYM, and \PLATON. 
In general, the alternative tolerances are 100 times the default tolerances, except in the case of \PLATON, where the default 
tolerances are divided by 100:
\begin{itemize}
\item \SPGLIB: $\verb|symprec|=1\times10^{-3}$~\AA
\item \FINDSYM: $\epsilon_{\mathrm{lattice}}=1\times10^{-3}$~\AA, $\epsilon_{\mathrm{atomic~position}}=1\times10^{-1}$~\AA
\item \PLATON: $\epsilon_{\mathrm{metric}}=0.01\degrees$, $\epsilon_{\mathrm{rotation}}=2.5\times10^{-3}$~\AA, 
  $\epsilon_{\mathrm{inversion}}=2.5\times10^{-3}$~\AA, $\epsilon_{\mathrm{translation}}=2.5\times10^{-3}$~\AA.
\end{itemize}
The results from the alternative tolerances are denoted with $^{+}$.

\subsection{Accuracy of space group analyses}
The \CIF\ files stored in the \ICSD\ contain information such as the structural parameters and 
atomic species/positions, as well as the space group (often reported from experiments), publication
date, and citation.
The experimentally reported space group information provides a unique validation 
opportunity for the various symmetry packages.
The mismatch counts between the reported and calculated space groups are shown in Table~\ref{table:symmetry_errors}.
The counts are additionally broken down by lattice and crystal system (Figure~\ref{fig:sg_inconsistencies}) 
to highlight the severity of the mismatch.
The full comparison of results is provided in the Supplementary Information, organized in tables
by the reported crystal system, with mismatches highlighted in red.

\AFLOWSYM\ shows the best agreement with the \ICSD\ with a deviation of about 1.5\% 
(reduced to 1.3\% if the mismatch is rectified at the loose tolerance).
The mismatch is almost halved when comparing only the lattice and crystal systems, 
suggesting the algorithm found similar/nearby space groups (\textit{e.g.}, see Figure~\ref{fig:tolerance_spectrum_scan}).
Using their respective default tolerances, \PLATON\ performs second best with a 5.6\% deviation, followed by \FINDSYM\ 
and \SPGLIB\ with deviations of about 6.6\% and 18.5\%, respectively. 
With the alternative tolerances, the overall number of mismatches decrease for each package: 
\PLATON\ reduces to 2.3\%, 
\FINDSYM\ reduces to 2.0\%, and
\SPGLIB\ reduces to 6.3\%.
Table~\ref{table:symmetry_errors} also shows that there are a number of systems for which \FINDSYM\ and \PLATON\
are unable to identify any space group.

Figure~\ref{fig:sg_inconsistencies} illustrates the space group mismatch from each package organized by lattice type.
Overall, \AFLOWSYM\ is the most consistent with the \ICSD\ for all lattice types for both the default and alternative 
tolerances, except for cubic systems where \FINDSYM\ has one less mismatch than \AFLOWSYM\ using the alternative tolerance.
The default tolerance values certainly play a role in the large deviation count, \textit{e.g.}, 
a tighter tolerance can yield a lower symmetry than expected.
This is evident with hexagonal and rhombohedral lattices, 
where \SPGLIB\ resolves isomorphic subgroups neglecting the 3-/6-fold rotations
(see Supplementary Information).
However, increases in tolerance do not necessarily yield more consistent space group determinations.  
Figure~\ref{fig:sg_inconsistencies} shows that the default tolerance is more accurate than the alternative tolerance 
for the triclinic (tri) and body-centered tetragonal (bct) systems calculated by \SPGLIB\ and \FINDSYM, respectively.
To guarantee consistent symmetry results, users of \SPGLIB, \FINDSYM, and \PLATON\ should 
tune the tolerance for each system.
The structure-specific tolerance choice and adaptive tolerance scheme incorporated into \AFLOWSYM\ allows for the automatic 
calculation of results that are generally consistent with experiments.  

Overall, the results indicate the strength of the \AFLOWSYM\ approach. 
Other packages can reach similar performance of \AFLOWSYM, but they require continuous {\it ad-hoc} user adjustments of tolerances, 
possibly producing results incommensurate with other characteristics of the systems, such as its Pearson symbol.
Only the self-consistent approach of \AFLOWSYM\ is ripe for the automation required by autonomous materials design.

\renewcommand{\arraystretch}{1.5}
\begin{table*}
  \centering
  \begin{tabular}{||l||c|c|c|c||}
    \hline
    \multicolumn{1}{||c||}{symmetry} & \multicolumn{1}{c|}{\AFLOWSYM} & \multicolumn{1}{c|}{\SPGLIB} & \multicolumn{1}{c|}{\FINDSYM} & \multicolumn{1}{c||}{\PLATON} \\
    \hline
    \hline
    lattice & \AFLOWCHECK & & & \\
    \hline
    superlattice & \AFLOWCHECK & & & \AFLOWCHECK\ (\verb|EQUAL|)\\
    \hline
    reciprocal lattice & \AFLOWCHECK & & & \\
    \hline
    crystal & \AFLOWCHECK & \AFLOWCHECK & \AFLOWCHECK & \AFLOWCHECK \\
    \hline
    crystal-spin & \AFLOWCHECK & \AFLOWCHECK & \AFLOWCHECK & \\
    \hline
  \end{tabular}
  \caption{\textbf{List of the symmetry descriptions provided by each of the four packages.}  The superlattice analysis refers to
    the structure symmetry if each atomic site is decorated equally (same atom type), while crystal-spin indicates the 
    structure symmetry including the magnetic moment of each atom.}
  \label{table:sym_features}
\end{table*}
\renewcommand{\arraystretch}{1.0}

\subsection{Symmetry characterizations and representations}
Of primary concern among the various standard packages is the identification and characterization
of crystal symmetry, \textit{i.e.}, a symmetry description considering the lattice and basis of atoms.
In addition, \AFLOWSYM\ characterizes crystals with a sequence of
symmetry-breaking features, including the lattice, superlattice (lattice with a uniform basis), 
crystal, and crystal-spin.
With the progression of symmetry-breaking, each characterization offers a 
new dimension of physical insight, and is of particular importance for 
understanding complex phenomena~\cite{Matano_NPhys_2016}.
The suite of characterizations\footnote{Some packages provide more information than listed in Table~\ref{table:sym_features}.
  For example, \PLATON\ presents additional useful structural/chemical information 
  such as bonding, coordination, planes, and torsions.
  However, the comparison presented in Table~\ref{table:sym_features} is limited 
  to symmetry information pertaining to space groups.}
offered by each package is presented in Table~\ref{table:sym_features}.  
{{With integration into the automated framework AFLOW, 
new tools and symmetry descriptions will continue to be incorporated.  
The forums at \url{aflow.org/forum} are the venues for presenting updates and discussing new functionalities.  
Anticipated future work includes going beyond translationally invariant structures and characterizing 
disordered/off-stoichiometric structures~\cite{curtarolo:art110,curtarolo:art112}.}}

Furthermore, \AFLOWSYM\ presents the symmetry operations in a wealth of representations.
Both \AFLOWSYM\ and \SPGLIB\ explicitly offer representations for the symmetry operations.\footnote{\FINDSYM\ and \PLATON\ do
provide the general Wyckoff position, though they do not explicitly present the symmetry operators.}
Table~\ref{table:symmetry_ops} compares the operation representations provided by the two packages.
Both provide the unit cell symmetry operators (coset representatives of the factor group).
\AFLOWSYM\ offers the symmetry operations in the rotation matrix (cartesian and fractional), axis-angle, generator, and
quaternion representations~\cite{karney_jmgm_2007,fritzer_saams_2001};
while \SPGLIB\ only provides the rotation matrix representation in its fractional form.
\AFLOWSYM\ also presents the corresponding mappings for each symmetry operation, almost entirely
eliminating the need to reapply the operators for symmetry-reduced analyses such as calculating
the force constants~\cite{curtarolo:art67,curtarolo:art125}.
Along with the factor group coset representatives, \AFLOWSYM\ provides the lattice point group,
reciprocal lattice point group, crystal point group, dual of the crystal point group, site point group, and space group symmetry operators.
Catering to electronic structure calculations, \AFLOWSYM\ also returns additional symmetry information
not explicitly provided by other routines, such as the 
Pearson symbol, Bravais lattice type, and Bravais lattice variation, necessary
for constructing the most efficient Brillouin zone~\cite{curtarolo:art58}.
The full set of descriptions and representations offered by \AFLOWSYM\ is detailed in Appendix~\ref{subsec:mathrepresentation}.

\renewcommand{\arraystretch}{1.5}
\begin{table*}
  \centering
  \begin{tabular}{||l||c|c||}
    \hline
    \multicolumn{1}{||c||}{operator information} & \multicolumn{1}{c|}{\AFLOWSYM} & \multicolumn{1}{c||}{\SPGLIB} \\
    \hline
    \hline
    operator type  & \AFLOWCHECK &  \\
    \hline
    Hermann-Mauguin  & \AFLOWCHECK &  \\
    \hline
    Sch{\"o}nflies  & \AFLOWCHECK & \\
    \hline
    transformation matrix (cartesian)  & \AFLOWCHECK&   \\
    \hline
    transformation matrix (fractional)  & \AFLOWCHECK& \AFLOWCHECK\\
    \hline
    generator matrix & \AFLOWCHECK&   \\
    \hline
    {{so(3) coefficients ($\mathbf{L}_{x},\mathbf{L}_{y},\mathbf{L}_{z}$)}} & \AFLOWCHECK&   \\
    \hline
    angle & \AFLOWCHECK& \\
    \hline
    axis & \AFLOWCHECK&   \\
    \hline
    quaternion (vector) & \AFLOWCHECK&   \\
    \hline
    {{quaternion ($2\times2$ matrix)}} & \AFLOWCHECK&   \\
    \hline
    {{quaternion ($4\times4$ matrix)}} & \AFLOWCHECK&   \\
    \hline
    {{su(2) coefficients (Pauli)}} & \AFLOWCHECK&   \\
    \hline
    inversion boolean & \AFLOWCHECK&   \\
    \hline
    internal translation (cartesian) & \AFLOWCHECK&   \\
    \hline
    internal translation (fractional) & \AFLOWCHECK& \AFLOWCHECK\\
    \hline
    atom index map & \AFLOWCHECK&   \\
    \hline
    atom type map & \AFLOWCHECK&   \\
    \hline
    lattice translation (cartesian) & \AFLOWCHECK & \\
    \hline
    lattice translation (fractional) & \AFLOWCHECK & \\
    \hline
  \end{tabular}
  \caption{\textbf{List of operation representations provided by \AFLOWSYM\ compared to \SPGLIB.}  The internal translations
    are only applicable for the coset representative of the factor group and space group symmetry operators.  Likewise, 
    the lattice translations are only applicable for the space group symmetry operators.}
  \label{table:symmetry_ops}
\end{table*}
\renewcommand{\arraystretch}{1.0}

\section{Using \AFLOWSYM}

\subsection{Input/output formats}
{{\AFLOWSYM\ reads crystal structure information from a geometry file 
containing the lattice vectors and atomic coordinates (coordinate model), 
which is treated as the \textit{bona fide} representation of the structure.  
Information can be lost during the transcription of the X-ray diffraction/reflection 
data to the coordinate model, resulting in a lower symmetry profile. 
While a means to verify the two representations offers higher fidelity symmetry 
descriptions, the diffraction data is not nearly as accessible as the coordinate model representation.  
Furthermore, the geometry file is the \textit{de facto} input format for \textit{ab-initio} packages, and 
thus \AFLOWSYM\ resolves the material's symmetry based on this representation.}} 

With \AFLOWSYM\ well-integrated into the high-throughput \textit{ab-initio} software package \AFLOW,
it can process many standard input file types, including
that of the \ICSD/\CSD~\cite{ICSD,ICSD3,Groom_CSD_2016} (\CIF),
\VASP~\cite{kresse_vasp,VASP4_2,vasp_cms1996,vasp_prb1996} (\POSCAR),
\QUANTUMESPRESSO~\cite{quantum_espresso_2009},
\ABINIT~\cite{gonze:abinit}, and \FHIAIMS~\cite{Blum_CPC2009_AIM}.

Furthermore, all symmetry functions support the JSON object output format. 
This allows \AFLOWSYM\ to be employed from other programming languages such as 
Java, Go, Ruby, Julia and Python; facilitating smooth integration  
into numerous applications and workflows~\cite{aflowPI,nomad}.
These functionalities can be accessed by either the command line or
a Python environment.  
A summary of the output for each command is provided in Appendix~\ref{subsec:outputlist}.

\subsection{Command line options} 
There are three main functions that provide all symmetry information for a given input structure.
These functions allow an optional tolerance value (\verb|tol|) to be specified via a number or the
strings ``tight'' or ``loose'' corresponding to $\epsilon_{\mathrm{tight}}$ and $\epsilon_{\mathrm{loose}}$, respectively.
To perform the symmetry analysis of a crystal,
the functions are called with the following commands:
\begin{widetext}
\begin{myitemize}
\item \verb!aflow --aflowSYM[=<tol>] [--print=txt|json] < file!
  \begin{myitemize}
  \item{Calculates and returns the symmetry operations for the lattice point group, reciprocal lattice point group, 
      coset representatives of the factor group, crystal point group, dual of the crystal point group, site symmetry, and space group.  It also returns the unique and equivalent
      sets of atoms.}
  \end{myitemize}
\item \verb!aflow --edata[=<tol>] [--print=txt|json] < file!
  \begin{myitemize}
  \item{Calculates and returns the extended crystallographic symmetry data (crystal, lattice,
      reciprocal lattice, and superlattice symmetry).}
  \end{myitemize}
\item \verb!aflow --sgdata[=<tol>] [--print=txt|json] < file!
  \begin{myitemize}
  \item{Calculates and returns the space group symmetry of the crystal.}
  \end{myitemize}
\end{myitemize}
\end{widetext}
Square brackets \verb|[...]| indicate optional arguments.
The \verb|--print| flag specifies the output format.  
The \verb|--aflowSYM| function stores the isometries of the different symmetry groups to their own files 
\verb|aflow.<group>.out| or \verb|aflow.<group>.json|.
The \verb|<group>| labels are as follows: \verb|pgroup| (lattice point group), 
\verb|pgroupk| (reciprocal lattice point group), \verb|fgroup| (coset representatives of the factor group), 
\verb|pgroup_xtal| (crystal point group), \verb|pgroupk_xtal|, (dual of the crystal point group), 
\verb|agroup| (site symmetry), and \verb|sgroup| (space group).

Crystal-spin symmetry functionality is also available in \AFLOWSYM.
The magnetic moment of each site 
{{(collinear or non-collinear)}}
can be specified for each of the commands
listed above by adding the magnetic moment flag: \verb![--magmom=m1,m2,...|INCAR|OUTCAR]!.
The magnetic moment information can be specified in three formats:
{\bf i.} explicitly via $m_{1}$, $m_{2}$, ... $m_{n}$ in the same order as the input file 
{{(or $m_{1,x}$, $m_{1,y}$, $m_{1,z}$, $m_{2,x}$ ... $m_{n,z}$ for non-collinear)}},
{\bf ii.} read from the \VASP\ \INCAR\ or
{\bf iii.} the \VASP\ \OUTCAR. 
Magnetic moment readers for other {\it{ab-initio}} codes will be added in later versions.

\subsection{Python environment}
\label{subsec:python_environment}
Given the recent prevalence of Python programming, we offer a module that employs \AFLOWSYM\ 
within a Python environment (see Appendix~\ref{subsec:python_module}).
It connects to a local \AFLOW\ installation and imports the \AFLOWSYM\ results into a 
\verb|Symmetry| class. 
A \verb|Symmetry| object is initialized with:

\begin{python}
  from aflow_sym import Symmetry
  from pprint import pprint

  with open('test.poscar', 'r') as input_file:
    sym = Symmetry(aflow_executable='./aflow')
    output = sym.get_edata(input_file)
    pprint(output)
\end{python}

\noindent By default, the \verb|Symmetry| object searches for an \AFLOW\ executable in
the \verb|PATH|. However, the location of an \AFLOW\ executable can be specified as
follows:

\verb|Symmetry(aflow_executable=`your_executable')|.

\noindent The symmetry object has three built-in methods, which correspond to the command line calls mentioned previously:

\begin{myitemize}
\item \verb|get_symmetry(input_file, tol, magmoms)|
\item \verb|get_edata(input_file, tol, magmoms)|
\item \verb|get_sgdata(input_file, tol, magmoms)|
\end{myitemize}
Each method requires a Python file handler (\verb|input_file|), while the tolerance (\verb|tol|) and magnetic moments of each site 
(\verb|magmoms|) are optional arguments.

\subsection{\AFLOWSYM\ support}
Functionality requests and bug reports should be posted on the \AFLOW\ Forum
\url{aflow.org/forum} under the board ``Symmetry analysis''.

\section{Conclusion}

In this article, we present \AFLOWSYM, a symmetry platform catered to --- but not limited to --- high-throughput
frameworks.  
We address problems stemming from numerical tolerance in symmetry analyses by using a mapping procedure uniquely designed 
to handle skewed cells and an advanced adaptive tolerance scheme.  
\AFLOWSYM\ also includes consistency checks of calculated isometries with respect to symmetry principles.
These solutions are validated against the experimental structures data reported by the \ICSD.
Comparison with other symmetry analysis suites, \SPGLIB, \FINDSYM, and \PLATON, shows that \AFLOWSYM\ is the most consistent with 
the \ICSD.

For general use of \AFLOWSYM, the routines include both a
standard text output and a JSON output for easy integration into other computational workflows.  Lastly, a
comprehensive list of the symmetry descriptions are presented (see Appendix~\ref{subsec:outputlist}), illustrating the vast symmetry 
information available to users of \AFLOWSYM.

\section{Acknowledgments}
DH acknowledges support from the Department of Defense through the
National Defense Science and Engineering Graduate (NDSEG) Fellowship Program.
CO acknowledges support from the National Science Foundation Graduate Research
Fellowship under Grant No. DGF1106401.
SC acknowledges support by DOD-ONR (N00014-17-1-2090) and
the Alexander von Humboldt-Foundation.

%
%
\appendix

\section{Crystallographic symmetry}

\label{sec:symmetry_groups}
\subsection{Mathematical group} 
A group is an abstract mathematical structure comprised of a set of elements ($g$), and an operation that combines 
two elements to form a third~\cite{Tinkham_GroupTheory_1992}.
There are four axioms that a group satisfies.
\begin{enumerate}
\item{closure: The combination of two elements with the operator yields an element that exists in the set; it does not create a new element outside the set.}
\item{associativity: The order of combining elements with the operator is inconsequential given the sequence of operands is unaltered.}
\item{identity: There exists a neutral element ($I$) that when combined with another element leaves that element unchanged ($gI=g$).}
\item{inverse: For each element $g$ in the set, there exists a corresponding inverse element $g^{-1}$, such that $gg^{-1}=I$.}
\end{enumerate}
An abelian group includes the additional axiom of commutativity.
These rules are the foundation of group theory and underline the construction of the different symmetry groups.

\subsection{Point group} 
A point group is a set of symmetry transformations about a fixed point $\left\{\mathbf{U}_{1}, \mathbf{U}_{2}, \dots, \mathbf{U}_{n}\right\}$ 
that leave a system invariant\footnote{Here, the rotation matrix $\mathbf{U}$ is used to represent the different 
symmetry groups; however, all group elements can also be described in 
axis-angle, matrix generator, and quaternion form.}.
The elements of the group are classified as
{\bf i.} $n$-fold rotations, where $n$ describes the rotation order ({\it i.e.}, the number of symmetric points it generates), 
{\bf ii.} inversions, and
{\bf iii.} roto-inversions --- compound operations comprised of a rotation and inversion.
Three dimensional crystals are confined to one of 32 point groups due to the 
crystallographic restriction theorem, which limits the rotation order in a periodic system to two-, three-, four-, 
and six-fold~\cite{tables_crystallography}.
The 32 crystallographic point groups are categorized into one of seven crystal systems: cubic, hexagonal, trigonal,
tetragonal, orthorhombic, monoclinic, and triclinic.  
The classifications are based on the lattice parameters
($a$, $b$, $c$, $\alpha$, $\beta$, $\gamma$) of the crystal.
\begin{itemize}
\item{\textbf{cubic:} $a=b=c$, $\alpha = \beta = \gamma = 90^{\circ}$}
\item{\textbf{hexagonal/trigonal:} $a=b\neq c$, $\alpha = \beta = 90^{\circ}, \gamma = 120^{\circ}$}
\item{\textbf{tetragonal:} $a=b\neq c$, $\alpha = \beta = \gamma = 90^{\circ}$}
\item{\textbf{orthorhombic:} $a\neq b\neq c$, $\alpha = \beta = \gamma = 90^{\circ}$}
\item{\textbf{monoclinic:} $a\neq b\neq c$, $\alpha = \gamma = 90^{\circ}, \beta \neq 90^{\circ}$}
\item{\textbf{triclinic:} $a\neq b\neq c$, $\alpha \neq \beta \neq \gamma \neq 90^{\circ}$}
\end{itemize}
In crystallography, two types of point groups are of particular importance --- the lattice and crystal (vector) point group.
Each operates in a different space: 
the lattice point group characterizes the symmetry of the lattice points (an affine space), 
while the crystal point group additionally considers the atomic basis and acts on 
the underlying vector space of the crystal face normals.
Fundamentally, the vector space captures the symmetry of the macroscopic crystal~\cite{tables_crystallography}.
The crystal point group operations are defined as the linear mappings of the vector space, \textit{i.e.},
the unique set of fixed-point transformations of the 
factor group\footnote{Without the relevant internal translations (complete coset representatives), the crystal point group operations
do not generally apply in the affine point space (lattice points and atoms), as is the case for non-symmorphic space groups.
Conversely, the set of operations that do apply in the 
point space define the site symmetries.}~\cite{tables_crystallography,Nespolo_pointgroups_2009}.
Owing to symmetry breaking from the atomic basis, the cardinality of the crystal point group
is at most as large as that of the lattice.
Furthermore, the dual (reciprocal) counterparts of the 
lattice and crystal point group play an important role in electronic structure theory:
resolving the symmetries of the Brillouin and irreducible Brillouin zones, respectively.
In \AFLOWSYM, the output for the lattice, reciprocal lattice, crystal, and dual of the crystal point group 
operations are labeled \verb|pgroup|, \verb|pgroupk|, \verb|pgroup_xtal|, and \verb|pgroupk_xtal|, respectively.

\subsection{Space group} 
In periodic systems, translational symmetry gives rise to another mathematical group --- the space group.  
Its elements are comprised of those found in the point group, along with glide (mirror and translation)
and screw (rotation and translation) operations.  
The translational degree of freedom extends the number of unique sets of symmetry operations to 230.  
The translations of a crystal are divided into lattice translations ($\mathbf{T}$) and internal translations ($\mathbf{t}$).
\begin{equation}
  \label{space_group}
  \left\{\mathbf{U}_{1}, \mathbf{U}_{2}, \dots, \mathbf{U}_{n} | \mathbf{T} + \mathbf{t}\right\}
\end{equation}
Subsequently, a space group describes the full symmetry of a periodic system.
The space group operations are labeled \verb|sgroup| in \AFLOWSYM.

\subsection{Factor group} 
\label{subsec:fgroup}
From the space group, the elements of the factor group are defined as the cosets of 
the subgroup of lattice translations ($\mathbf{T}$): 
\begin{align}
  \label{factor_group}
  \left\{\mathbf{I} | \mathbf{0}\right\} \left\{\mathbf{I}|\mathbf{T}\right\}, \, \, 
  \left\{\mathbf{U}_{i} | \mathbf{t}_{i}\right\} \left\{\mathbf{I}|\mathbf{T}\right\}, \, \,
  \left\{\mathbf{U}_{j} | \mathbf{t}_{j}\right\} \left\{\mathbf{I},\mathbf{T}\right\}, \, \, \dots \, ,
\end{align}
where $\mathbf{U}_{i}$ are the point group operations, $\mathbf{t}_{i}$ are the associated internal translation, 
and $\mathbf{I}$ is the identity.
The unit cell symmetry is exposed via the coset representatives:
\begin{align}
  \label{coset_representatives}
  &\left\{\mathbf{I} | \mathbf{0}\right\}, \,
    \left\{\mathbf{U}_{i} | \mathbf{t}_{i}\right\}, \,
    \left\{\mathbf{U}_{j} | \mathbf{t}_{j}\right\}, \, \dots \, .
\end{align}
The coset representatives themselves do not necessarily form a mathematical group, since they violate the closure condition.  
Repeated application of an internal translation will eventually traverse beyond the unit cell.
The unit cell symmetry elements (coset representatives) are labeled \verb|fgroup| in \AFLOWSYM.

In general, there exists a homomorphism between the factor group and
the crystal point group, {\it i.e.}, the factor group cardinality is an integer
multiple of the crystal point group cardinality.  
The multiplicative factor ($m$) is dictated by the number of internal translations in the system.  
A crystal in a primitive representation exhibits an isomorphic correspondence
($m=1$), while non-primitive representations possess the general homomorphic relationship ($m>1$).

\subsection{Site point group} 
The site point group --- or site symmetry --- describes the point group symmetry
centered on a single site in the crystal, revealing the local symmetry environment.
The analysis is performed on each atomic site in the crystal, with symmetrically 
equivalent atoms (Wyckoff positions) exhibiting the same point group symmetries.  
The origin of the fixed-point operations differentiates the site symmetry from the lattice/crystal 
point group, which are centered on the unit cell origin.
In the finite difference method for calculating phonons, the unique distortions for a given atomic site are resolved 
with its site symmetry~\cite{curtarolo:art67,curtarolo:art125}.
In \AFLOWSYM, the site symmetry elements are designated by \verb|agroup| (``atomic site group'').

\subsection{Crystal-spin symmetry} 
Introducing the spin degree of freedom can break crystal symmetry.
\AFLOWSYM\ includes functionality for a crystal-spin (lattice, atoms, and spin) description, 
including the relevant point group, factor group, space group, and site symmetry operations.
For magnetic systems, these are the symmetry descriptions employed by {\it{ab-initio}} packages, such as 
\VASP~\cite{kresse_vasp,VASP4_2,vasp_cms1996,vasp_prb1996}.
Note that the crystal-spin symmetry differs from the magnetic symmetry, which accounts for time-reversal symmetry
(spin-flips).
The magnetic symmetry will be incorporated into \AFLOWSYM\ in a later version.

\section{Mathematical representation of symmetry}

\label{subsec:mathrepresentation}
\indent Symmetry elements are characterized into three types of transformation: translation, fixed-point, and fixed-point-free 
(a combination of the two, {\it i.e.}, screw and glide operations)~\cite{tables_crystallography}.  
A translation is generally indicated by $3\times1$ vectors
\begin{equation}
  \label{translation}
  \mathbf{t} = \left(
    \begin{array}{c}
      t_{1} \\
      t_{2} \\
      t_{3} \\
    \end{array}
  \right).
\end{equation}
Fixed-point symmetries O(3) describing rotations, inversions, and roto-inversions are
represented by rotation matrices.
The rotation symmetries SO(3), {\it i.e.}, a subgroup of the orthogonal group O(3), can be represented in 
three additional forms: axis-angle, matrix generator, and quaternions.
\AFLOWSYM\ provides the symmetry operations for rotations in each of these four forms, which are discrete 
subgroups of the continuous SO(3) group.

\subsection{Rotation matrix} 
A rotation matrix describes a transformation between two reference frames.
In three-dimensions, the symmetry operators are $3\times3$ square matrices with the following form
\begin{equation}
  \label{rotation}
  \mathbf{U} =
  \left(
    \begin{array}{ccc}
      u_{11} \, & u_{12} \, & u_{13} \\
      u_{21} \, & u_{22} \, & u_{23} \\
      u_{31} \, & u_{32} \, & u_{33}
    \end{array}
  \right).
\end{equation}
All transformations are unitary (norm preserving), and therefore have $\mathrm{det}\left(\mathbf{U}\right)=\pm 1$.
The matrix representation affords fast computation through use of optimized linear algebra computational
packages.

\subsection{Axis-angle} 
Rotation operations are also characterized by their axis and angle of rotation.
The axis, $\mathbf{\hat{r}}=(r_1, r_2, r_3)$, indicates the direction of the rotation operator,
pointing perpendicular to the fixed point motion.
The angle, $\theta$, specifies the magnitude of the rotational motion (following the right-hand rule). 
The angle and axis components are related to the matrix elements of \textbf{U} by
\begin{equation}
  \begin{aligned}
    \theta&=\textrm{cos}^{-1}\left(\dfrac{\textrm{Tr}(\textbf{U})-1}{2}\right), \\
    r_d&=\sqrt[]{(u_{32}-u_{23})^2+(u_{13}-u_{31})^2+(u_{21}-u_{12})^2}, \\
    r_1&=\dfrac{u_{32}-u_{23}}{r_d}, \, r_2=\dfrac{u_{13}-u_{31}}{r_d}, \, r_3=\dfrac{u_{21}-u_{12}}{r_d},
  \end{aligned}
\end{equation}
where $\textrm{Tr}(\textbf{U})$ is the trace of $\textbf{U}$.
The axis-angle representation is directly applied to a point $\mathbf{p}$ via Rodrigues' rotation 
formula
\begin{equation}
  \mathbf{p}_{\mathrm{rot}} = \mathbf{p}\cos\theta + \left(\mathbf{\hat{r}}\times\mathbf{p}\right) \sin\theta + 
                              \mathbf{\hat{r}} \left(\mathbf{\hat{r}}\cdot\mathbf{p}\right)\left(1-\cos\theta\right),
\end{equation}
where $\mathbf{p}_{\mathrm{rot}}$ is the rotated point. 
This description highlights the operation order $n$ via $n = 360^{\circ}/\theta$ and identifies the
conventional cell lattice vectors, since they are parallel to certain symmetry axes.

\subsection{Matrix generator}
The Lie group SO(3) grants the use of the corresponding Lie algebra so(3), which are comprised of 
the infinitesimal matrix generators $\mathbf{G}$.
The generator is a skew-symmetric matrix that describes the rotation about a symmetry axis, with the following form:
\begin{equation}
  \label{generator}
  \mathbf{G} =
  \left(
    \begin{array}{ccc}
      0 \, & -r_{3} \, & r_{2} \\
      r_{3} \, & 0 \, & -r_{1} \\
      -r_{2} \, & r_{1} \, & 0
    \end{array}
  \right),
\end{equation}
where $r_{1}$, $r_{2}$, $r_{3}$ are the components of the symmetry unit axis $\mathbf{\hat{r}}$.
The identity and inverse elements have no axis; therefore, the generator is not defined and is returned as a zero matrix.
While the rotation matrix transforms one reference frame to another, the 
generator operates about a single axis.
The matrix exponential of the generator with the angle maps the operations into the rotation matrix form
($\mathbf{U}=\exp(\theta\mathbf{G})$).
For convenience, \AFLOWSYM\ returns the generator multiplied with the angle $\mathbf{A}=\theta\mathbf{G}$.
{{\AFLOWSYM\ also provides the expansion coefficients of the generator matrix onto the following so(3) basis:
\begin{equation}
  \mathbf{G} = x\mathbf{L}_{x} + y\mathbf{L}_{y} + z\mathbf{L}_{z}
\end{equation}
where
\begin{equation}
  \begin{aligned}
  \mathbf{L}_{x} &= \left(
    \begin{array}{ccc}
      0 \, & 0 \, & 0 \\
      0 \, & 0 \, & -1 \\
      0 \, & 1 \, & 0 \\
    \end{array}
  \right) \, , \,
  \mathbf{L}_{y} = \left(
    \begin{array}{ccc}
      0 \, & 0 \, & 1 \\
      0 \, & 0 \, & 0 \\
      -1 \, & 0 \, & 0 \\
    \end{array}
  \right) \\
  \mathbf{L}_{z} &= \left(
    \begin{array}{ccc}
      0 \, & -1 \, & 0 \\
      1 \, & 0 \, & 0 \\
      0 \, & 0 \, & 0 \\
    \end{array}
  \right).
  \end{aligned}
\end{equation}
The expansion coefficients $x$, $y$, and $z$ of this basis set are the unit axis components $r_{1}$, $r_{2}$, and $r_{3}$, respectively.
}}

\subsection{Quaternion} 
A quaternion is a mathematical representation of 3D space with both real and imaginary components.
Though developed in 1843, the quaternion has only recently gained relevance through the field of computer
graphics and modeling. 
As opposed to using a nine-element $3 \times 3$ matrix to represent a rotation in space,
quaternions have a concise format consisting of four components.
The reduced element count increases computational efficiency, and thus is particularly
suitable for high-throughput frameworks.

Given an axis and angle, the corresponding quaternion
representation, $\mathbf{q} = (q_0, q_1, q_2, q_3)$, is
\begin{equation}
  \begin{aligned}
    q_0 &= \cos(\theta/2), \\
    q_1 &= r_1\sin(\theta/2), \\
    q_2 &= r_2\sin(\theta/2), \\
    q_3 &= r_3\sin(\theta/2),
  \end{aligned}
\end{equation}
{{which are equivalent to the Euler parameters.
Alternate forms of the quaternion are $2\times2$ and $4\times4$ matrices. 
The complex $2\times2$ unitary matrix of a quaternion is
\begin{equation}
  \mathbf{C} = \left(
    \begin{array}{cc}
      q_{0} + q_{3}i \, & q_{2} + q_{1}i \\
      -q_{2} + q_{1}i \, &  q_{0} - q_{3}i 
    \end{array}
  \right),
\end{equation}
which is an element of the SU(2) Lie group.
The $\mathbf{C}$ matrix can be expanded onto a basis formed by the Pauli matrices:
\begin{equation}
  \boldsymbol{\sigma}_{1} = \left(
    \begin{array}{cc}
      0 \, & 1 \\
      1 \, & 0
    \end{array}
  \right), \,
  \boldsymbol{\sigma}_{2} = \left(
    \begin{array}{cc}
      0 \, & -i \\
      i \, & 0
    \end{array}
  \right), \,
  \boldsymbol{\sigma}_{3} = \left(
    \begin{array}{cc}
      1 \, & 0 \\
      0 \, & -1
    \end{array}
  \right), 
\end{equation}
where multiplying by $i$ (=$\sqrt{-1}$) yields the following decomposition:
\begin{equation}
  \mathbf{C} = q_{0}\mathbf{I} + q_{1}i\boldsymbol{\sigma}_{1} + q_{2}i\boldsymbol{\sigma}_{2} + q_{3}i\boldsymbol{\sigma}_{3}.
\end{equation}
The corresponding Lie algebra, su(2), is~\cite{Gilmore_LieGroups_2008} 
\begin{equation}
  \mathbf{g} = \frac{i}{2}\left(
  \begin{array}{cc}
  r_{3} \, & r_{1}-r_{2}i \\
  r_{1}+r_{2}i \, & -r_{3}
  \end{array}
  \right).
\end{equation}
\AFLOWSYM\ lists the su(2) generator coefficients expanded on the Pauli matrices
\begin{equation}
  \mathbf{g} = x\boldsymbol{\sigma}_{1} + y\boldsymbol{\sigma}_{2} + z\boldsymbol{\sigma}_{3},
\end{equation}
where the expansion coefficients $x$, $y$, and $z$ are $(i/2)r_{1}$, $(i/2)r_{2}$, and $(i/2)r_{3}$, respectively.
Similar to the SO(3) rotations, the matrix exponential of the su(2) generator $\mathbf{g}$ with the angle maps the operations 
into the complex $2\times2$ SU(2) matrix ($\mathbf{C}=\exp(\theta\mathbf{g})$).
The $4\times4$ matrix representation of the quaternion is\\
}}
\begin{equation}
  \textbf{Q}=
  \left(\begin{array}{cccc} q_0 & q_1 & q_2 & q_3\\ -q_1 & q_0 & -q_3& q_2\\
          -q_2 & q_3 & q_0 & -q_1 \\ -q_3& -q_2 & q_1 & q_0 \end{array}\right);
\end{equation}
which includes all four components of the quaternion vector in a matrix,
allowing transformations to be performed through matrix multiplication
rather than quaternion algebra. This method is useful for performing
operations with other transformations in matrix or vector form, whereas
the quaternion vector notation has its own algebra similar to the operations
between complex numbers with an additional scalar component ($q_0$).

\subsection{Basis transformations of operators} 
The representations of the symmetry operations are basis dependent and are
customarily given with respect to cartesian or fractional coordinates systems.
It is straight-forward to transform symmetry operations between these vector spaces via a basis change.
In matrix notation, the fixed-point operation in cartesian ($\mathbf{U}_{\mathrm{c}}$) and fractional 
($\mathbf{U}_{\mathrm{f}}$) coordinates are related via the following similarity transformations:
\begin{equation}
  \begin{aligned}
    \label{similarity_transforms}
    \mathbf{U}_{\mathrm{f}} &= \mathbf{L}^{-1}\mathbf{U}_{\mathrm{c}}\mathbf{L}, \\
    \mathbf{U}_{\mathrm{c}} &= \mathbf{L}\mathbf{U}_{\mathrm{f}}\mathbf{L}^{-1}. \\
  \end{aligned}
\end{equation}
Here, $\mathbf{L}$ is the column-space form of the lattice vectors:
\begin{equation}
  \mathbf{L} = \left(
    \begin{array}{ccc}
      \mathbf{a} \, & \mathbf{b} \, & \mathbf{c} \\
    \end{array}
  \right) = \left(
    \begin{array}{ccc}
      a_{1} \, & b_{1} \, & c_{1} \\
      a_{2} \, & b_{2} \, & c_{2} \\
      a_{3} \, & b_{3} \, & c_{3} \\
    \end{array}
  \right),
\end{equation}
where $a_{i}$, $b_{i}$, and $c_{i}$ are the corresponding components of the lattice vectors.
A translation vector $\mathbf{t}_{\mathrm{c(f)}}$ is transformed between cartesian and fractional coordinates by $\mathbf{t}_{\mathrm{f}} = \mathbf{L}^{-1}\mathbf{t}_{\mathrm{c}}$ and $\mathbf{t}_{\mathrm{c}} = \mathbf{L}\mathbf{t}_{\mathrm{f}}$.

\subsection{Example representations}
An example of a three-fold rotation in cartesian coordinates is shown below in its rotation matrix, axis-angle, matrix generator, and
quaternion vector and matrix representations.
\begin{equation*}
  \begin{aligned}
    \textbf{U}_{\textrm{3-fold}} &=
    \left(\begin{array}{ccc} 0 & \textrm{-}1 & 0\\ 0 & 0 & \textrm{-}1\\ 1 & 0 & 0 \end{array}\right) \\
    \mathbf{\hat{r}} &= (0.57735, -0.57735, 0.57735)\\
    \theta &= 120^{\circ}\\
    \mathbf{A} &= \left(\begin{array}{ccc} 0.0 & -1.2092 & -1.2092 \\ 1.2092 & 0.0 & -1.2092 \\
                          1.2092 & 1.2092 & 0.0 \end{array}\right)\\
    \mathbf{q} &= (0.5, 0.5, -0.5, 0.5)\\
    {{\mathbf{C}{}}} & {{{}= \left(\begin{array}{cc} 0.5 + 0.5i & -0.5 + 0.5i \\
                          0.5 + 0.5i & 0.5 - 0.5i \end{array}\right)
    }} \\
    \mathbf{Q} &= \left(\begin{array}{cccc} 0.5 & 0.5 & -0.5 & 0.5\\
                          -0.5 & 0.5 & -0.5 & -0.5\\ 0.5 & 0.5 & 0.5 & -0.5 \\ -0.5 & 0.5 & 0.5 & 0.5 \end{array}\right)
  \end{aligned}
\end{equation*}

\section{Extreme cases of minimal distance discrepancy between cartesian and fractional spaces}

\label{sec:metric_warping}
The bring-in-cell procedure applied to a crystal with lattice parameters $a=b=c=5~\mathrm{\AA}$,
$\alpha=\gamma=90^\circ$ and $\beta=60^\circ$ identifies the minimum distance between the fractional
coordinates $(0,0,1/2)$ and $(1/2,0,0)$ to be $||\mathbf{\widetilde{d}}^{\mathrm{min}}_{\mathrm{c}}||=4.3301~\mathrm{\AA}$, 
compared to the true minimum of $||\mathbf{d}^{\mathrm{min}}_{\mathrm{c}}||=2.5~\mathrm{\AA}$.
A more extreme mismatch occurs if $\beta=5^\circ$, yielding a minimum of 
$||\mathbf{\widetilde{d}}^{\mathrm{min}}_{\mathrm{c}}||=4.9952~\mathrm{\AA}$ with the bring-in-cell method, 
differing significantly from the true minimum of $||\mathbf{d}^{\mathrm{min}}_{\mathrm{c}}||=0.2181~\mathrm{\AA}$.
Applying the heuristic threshold to the aforementioned skewed examples give bounds of $\epsilon_{\mathrm{max}}=1.2130~\mathrm{\AA}$ 
(with $d^{\mathrm{nn(min)}}_{\mathrm{c}}=2.4259~\mathrm{\AA}$) 
and $\epsilon_{\mathrm{max}}=0.0017~\mathrm{\AA}$ (with $d^{\mathrm{nn(min)}}_{\mathrm{c}}=0.4362~\mathrm{\AA}$) 
for $\beta=60^\circ$ and $\beta=5^\circ$, respectively.
Both thresholds are sufficiently below the true minimum distances --- even in the worst cases --- validating our choice of the
heuristic threshold.

\section{AFLOW-SYM details}

\subsection{Python module}
\label{subsec:python_module}
The module to run the \AFLOWSYM\ commands referenced in Section \ref{subsec:python_environment} is provided below. 
\begin{python}
import json
import subprocess
import os

class Symmetry:

    def __init__(self, aflow_executable='aflow'):
        self.aflow_executable = aflow_executable

    def aflow_command(self, cmd):
        try:
            return subprocess.check_output(
                self.aflow_executable + cmd,
                shell=True
            )
        except subprocess.CalledProcessError:
            print "Error aflow executable not found at: " + self.aflow_executable

    def get_symmetry(self, input_file, tol=None, magmoms=None):
        fpath = os.path.realpath(input_file.name)
        command = ' --aflowSYM'
        output = ''

        if tol:
            command += '=' + str(tol)
        if magmoms:
            command += ' --magmom=' + magmoms

        output = self.aflow_command(
            command + ' --print=json --screen_only' + ' < ' + fpath
        )
        res_json = json.loads(output)
        return res_json

    def get_edata(self, input_file, tol=None, magmoms=None):
        fpath = os.path.realpath(input_file.name)
        command = ' --edata'
        output = ''

        if tol:
            command += '=' + str(tol)
        if magmoms:
            command += ' --magmom=' + magmoms

        output = self.aflow_command(
            command + ' --print=json' + ' < ' + fpath
        )
        res_json = json.loads(output)
        return res_json

    def get_sgdata(self, input_file, tol=None, magmoms=None):
        fpath = os.path.realpath(input_file.name)
        command = ' --sgdata'
        output = ''

        if tol:
            command += '=' + str(tol)
        if magmoms:
            command += ' --magmom=' + magmoms

        output = self.aflow_command(
            command + ' --print=json' + ' < ' + fpath
        )
        res_json = json.loads(output)
        return res_json
\end{python}

\subsection{Output list} 
\label{subsec:outputlist}
This section details the output fields for the symmetry group operations,
extended crystallographic data (\verb|edata|), and space group data (\verb|sgdata|) routines.
The lists describe the keywords as they appear in the JSON format.
Similar keywords are used for the standard text output.

\def\description{\item {{\it Description:}\ }}
\def\type{\item {{\it Type:}\ }}
\def\similarto{\item {{\it Similar to:}\ }}
\def\bluedescription{{{\item {{\it Description:}\ }}}} %
\def\bluetype{{{\item {{\it Type:}\ }}}}               %
\def\bluesimilarto{{{\item {{\it Similar to:}\ }}}}    %
\noindent\textbf{Symmetry operations output.}
\begin{myitemize}
\item \verb|pgroup|
  \begin{myitemize}
    \description lattice point group symmetry operations.
    \type \verb|array of symmetry operator objects|
  \end{myitemize}
\item \verb|pgroupk|
  \begin{myitemize}
    \description reciprocal lattice point group symmetry operations.
    \type \verb|array of symmetry operator objects|
  \end{myitemize}
\item \verb|fgroup|
  \begin{myitemize}
    \description coset representative of factor group symmetry operations.
    \type \verb|array of symmetry operator objects|
  \end{myitemize}
\item \verb|pgroup_xtal|
  \begin{myitemize}
    \description crystal point group symmetry operations.
    \type \verb|array of symmetry operator objects|
  \end{myitemize}
\item \verb|pgroupk_xtal|
  \begin{myitemize}
    \description dual of the crystal point group symmetry operations.
    \type \verb|array of symmetry operator objects|
  \end{myitemize}
\item \verb|sgroup|
  \begin{myitemize}
    \description space group symmetry operations out to a given radius.
    \type \verb|array of symmetry operator objects|
  \end{myitemize}
\item \verb|iatoms|
  \begin{myitemize}
    \description groupings of symmetrically equivalent/unique atoms.
    \type \verb|iatom object|
  \end{myitemize}
\item \verb|agroup|
  \begin{myitemize}
    \description site (atom) symmetry operations (point group).
    \type \verb|array of symmetry operator objects|
  \end{myitemize}
\end{myitemize}
\hfill\break
Each symmetry group contains an array of symmetry objects, including the operation representations
listed in Table~\ref{table:symmetry_ops}.  The \verb|symmetry operator| object contains the following:
\begin{myitemize}
\item \verb|Hermann_Mauguin|
  \begin{myitemize}
    \description Hermann-Mauguin symbol of the symmetry operation.
    \type \verb|string|
  \end{myitemize}
\item \verb|Schoenflies|
  \begin{myitemize}
    \description Sch{\"o}nflies symbol of the symmetry operation. 
    \type \verb|string|
  \end{myitemize}
\item \verb|Uc|
  \begin{myitemize}
    \description transformation matrix with respect to cartesian coordinates.
    \type $3 \times 3$ \verb|array|
  \end{myitemize}
\item \verb|Uf|
  \begin{myitemize}
    \description transformation matrix with respect to fractional coordinates.
    \type $3 \times 3$ \verb|array|
  \end{myitemize}
\item \verb|angle|
  \begin{myitemize}
    \description angle corresponding to symmetry operation.
    \type \verb|float|
  \end{myitemize}
\item \verb|axis|
  \begin{myitemize}
    \description axis of symmetry operation.
    \type $3 \times 1$ \verb|array|
  \end{myitemize}
\item \verb|generator|
  \begin{myitemize}
    \description matrix generator of symmetry operation.
    \type $3 \times 3$ \verb|array|
  \end{myitemize}
\item \cverb|generator_coefficients| %
  \begin{myitemize}
    \bluedescription {{matrix generator expansion coefficients onto $\mathbf{L}_{x}$, $\mathbf{L}_{y}$, and $\mathbf{L}_{z}$ basis.}}
    \bluetype {{$3 \times 1$}} \cverb|array| %
  \end{myitemize}
\item \verb|group|
  \begin{myitemize}
    \description specifies the group type (``\verb|pgroup|'', ``\verb|pgroupk|``, ``\verb|fgroup|'', ``\verb|pgroup_xtal|'',
    ``\verb|pgroupk_xtal|, ``\verb|sgroup|'', and ``\verb|agroup|'').
    \type \verb|string|
  \end{myitemize}
\item \verb|inversion|
  \begin{myitemize}
    \description indicates if inversion exists.
    \type \verb|bool|
  \end{myitemize}
\item \verb|quaternion_matrix|
  \begin{myitemize}
    \description quaternion matrix.
    \type $4 \times 4$ \verb|array|
  \end{myitemize}
\item \cverb|SU2_matrix| %
  \begin{myitemize}
    \bluedescription {{complex quaternion matrix; element of SU(2).}}
    \bluetype {{$2 \times 2$}} \cverb|array| %
  \end{myitemize}
\item \cverb|su2_coefficients| %
  \begin{myitemize}
    \bluedescription {{su(2) generator coefficients onto Pauli matrices ($\boldsymbol{\sigma}_{1}$, $\boldsymbol{\sigma}_{2}$, and $\boldsymbol{\sigma}_{3})$.}}
    \bluetype {{$3 \times 1$}} \cverb|array| %
  \end{myitemize}
\item \verb|quaternion_vector|
  \begin{myitemize}
    \description quaternion vector.
    \type $4 \times 1$ \verb|array|
  \end{myitemize}
\item \verb|type|
  \begin{myitemize}
    \description point group operation type (unity, rotation, inversion, or roto-inversion).
    \type \verb|string|
  \end{myitemize}
\item \verb|ctau|
  \begin{myitemize}
    \description internal translation component in cartesian coordinates (``\verb|fgroup|'' and ``\verb|sgroup|'' only).
    \type $3 \times 1$ \verb|array|
  \end{myitemize}
\item \verb|ftau|
  \begin{myitemize}
    \description internal translation component in fractional coordinates (``\verb|fgroup|'' and ``\verb|sgroup|'' only).
    \type $3 \times 1$ \verb|array|
  \end{myitemize}
\item \verb|ctrasl|
  \begin{myitemize}
    \description lattice translation component in cartesian coordinates (``\verb|sgroup|'' only).
    \type $3 \times 1$ \verb|array|
  \end{myitemize}
\item \verb|ftrasl|
  \begin{myitemize}
    \description lattice translation component in fractional coordinates (``\verb|sgroup|'' only).
    \type $3 \times 1$ \verb|array|
  \end{myitemize}
\end{myitemize}
\hfill\break
The \verb|iatom| object contains:
\begin{myitemize}
\item \verb|inequivalent_atoms|
  \begin{myitemize}
    \description symmetrically distinct atom indices.
    \type \verb|array|
  \end{myitemize}
\item \verb|equivalent_atoms|
  \begin{myitemize}
    \description groupings of symmetrically equivalent atom indices.
    \type \verb|2D array|
  \end{myitemize}
\end{myitemize}

\hfill\break
\noindent\textbf{edata output.}
\begin{myitemize}
\item \verb|lattice_parameters|
  \begin{myitemize}
    \description lattice parameters in units of Angstroms and degrees ($a,b,c,\alpha,\beta,\gamma$).
    \type $6 \times 1$ \verb|array|
    \similarto
    \begin{myitemize}
    \item \FINDSYM: \verb|Lattice parameters, a, b, c, alpha,| \\ \verb|beta, gamma:|
    \item \PLATON: first six fields in the line containing \verb|CELL|
    \end{myitemize}
  \end{myitemize}
\item \verb|lattice_parameters_Bohr_deg|
  \begin{myitemize}
    \description lattice parameters in units of Bohr and degrees ($a,b,c,\alpha,\beta,\gamma$).
    \type $6 \times 1$ \verb|array|
  \end{myitemize}
\item \verb|volume|
  \begin{myitemize}
    \description real space cell volume.
    \type \verb|float|
    \similarto
    \begin{myitemize}
    \item \PLATON: last field in the line containing \verb|CELL|.
    \end{myitemize}
  \end{myitemize}
\item \verb|c_over_a|
  \begin{myitemize}
    \description ratio of $c$ and $a$ lattice parameters.
    \type \verb|float|
  \end{myitemize}
\item \verb|Bravais_lattice_type|
  \begin{myitemize}
    \description Bravais lattice of the crystal (``FCC'', ``BCC'', ``CUB'', ``HEX'', ``RHL'', etc.).
    \type \verb|string|
  \end{myitemize}
\item \verb|Bravais_lattice_variation_type|
  \begin{myitemize}
    \description lattice variation type of the crystal in the \AFLOW\ standard~\cite{curtarolo:art58}.
    \type \verb|string|
  \end{myitemize}
\item \verb|Bravais_lattice_system|
  \begin{myitemize}
    \description Bravais lattice of the crystal.
    \type \verb|string|
    \similarto
    \begin{myitemize}
    \item \PLATON: \verb|CrystalSystem| column in \verb|Cell Lattice| table.
    \end{myitemize}
  \end{myitemize}
\item \verb|Pearson_symbol|
  \begin{myitemize}
    \description Pearson symbol of the crystal.
    \type \verb|string|
  \end{myitemize}
\item \verb|crystal_family|
  \begin{myitemize}
    \description crystal family.
    \type \verb|string|
  \end{myitemize}
\item \verb|crystal_system|
  \begin{myitemize}
    \description crystal system.
    \type \verb|string|
  \end{myitemize}
\item \verb|point_group_Hermann_Mauguin|
  \begin{myitemize}
    \description Hermann-Mauguin symbol corresponding to the point group of the crystal.
    \type \verb|string|
    \similarto
    \begin{myitemize}
    \item \SPGLIB: \verb|SpglibDataset.pointgroup_symbol|.
    \end{myitemize}
  \end{myitemize}
\item \verb|point_group_Schoenflies|
  \begin{myitemize}
    \description Sch{\"o}nflies symbol for the point group of the crystal.
    \type \verb|string|
  \end{myitemize}
\item \verb|point_group_orbifold|
  \begin{myitemize}
    \description orbifold of the point group.
    \type \verb|string|
  \end{myitemize}
\item \verb|point_group_type|
  \begin{myitemize}
    \description point group type of the crystal.
    \type \verb|string|
  \end{myitemize}
\item \verb|point_group_order|
  \begin{myitemize}
    \description number of point group operations describing the crystal.
    \type \verb|int|
  \end{myitemize}
\item \verb|point_group_structure|
  \begin{myitemize}
    \description point group structure of the crystal.
    \type \verb|string|
  \end{myitemize}
\item \verb|Laue|
  \begin{myitemize}
    \description Laue symbol of the crystal.
    \type \verb|string|
    \similarto
    \begin{myitemize}
    \item \PLATON: field after the line containing \verb|Laue|.
    \end{myitemize}
  \end{myitemize}
\item \verb|crystal_class|
  \begin{myitemize}
    \description crystal class.
    \type \verb|string|
  \end{myitemize}
\item \verb|space_group_number|
  \begin{myitemize}
    \description space group number.
    \type \verb|int|
    \similarto
    \begin{myitemize}
    \item \SPGLIB: \verb|SpglibDataset.spacegroup_number|.
    \item \FINDSYM: field after line containing \\\verb|_symmetry_Int_Tables_number|.
    \item \PLATON: field after line containing \verb|No| (number).
    \end{myitemize}
  \end{myitemize}
\item \verb|space_group_Hermann_Mauguin|
  \begin{myitemize}
    \description Hermann-Mauguin space group label.
    \type \verb|string|
    \similarto
    \begin{myitemize}
    \item \SPGLIB: \verb|SpglibDataset.International_symbol|.
    \item \FINDSYM: field after line containing \\\verb|_symmetry_space_group_name_H-M|.
    \item \PLATON: field after line containing \verb|Space Group  H-M|.
    \end{myitemize}
  \end{myitemize}
\item \verb|space_group_Hall|
  \begin{myitemize}
    \description Hall space group label.
    \type \verb|string|
    \similarto
    \begin{myitemize}
    \item \SPGLIB: \verb|SpglibDataset.hall_symbol|.
    \item \FINDSYM: field after line containing \\\verb|_space_group.reference_setting|.
    \item \PLATON: field after line containing \verb|Space group - Hall|.
    \end{myitemize}
  \end{myitemize}
\item \verb|space_group_Schoenflies|
  \begin{myitemize}
    \description Sch{\"o}nflies space group label.
    \type \verb|string|
    \similarto
    \begin{myitemize}
    \item \SPGLIB: \verb|Spg_get_schoenflies|.
    \item \FINDSYM: second field after line containing \verb|Space Group|.
    \item \PLATON: field after line containing \verb|Schoenflies|.
    \end{myitemize}
  \end{myitemize}
\item \verb|setting_ITC|
  \begin{myitemize}
    \description \ITC\ setting of conventional cell (\AFLOWSYM\ defaults to the first setting that appears in the \ITC\ and the hexagonal
    setting for rhombohedral systems).
    \type \verb|int|
    \similarto
    \begin{myitemize}
    \item \SPGLIB: \verb|SpglibDataset.choice|.
    \end{myitemize}
  \end{myitemize}
\item \verb|origin_ITC|
  \begin{myitemize}
    \description corresponding origin shift of the crystal to align with the \ITC\ representation.
    \type $3 \times 1$ \verb|array|
    \similarto
    \begin{myitemize}
    \item \SPGLIB: \verb|SpglibDataset.choice|.
    \item \FINDSYM: field after line containing \verb|Origin at|.
    \item \PLATON: field after line containing \verb|Origin Shifted to|.
    \end{myitemize}
  \end{myitemize}
\item \verb|general_position_ITC|
  \begin{myitemize}
    \description general Wyckoff position ($x,y,z$) as indicated by the \ITC.
    \type \verb|2D array|
    \similarto
    \begin{myitemize}
    \item \FINDSYM: field after line containing \\\verb|_space_group_symop_operation_xyz|.
    \item \PLATON: in the \verb|Symmetry Operation(s)| table.
    \end{myitemize}
  \end{myitemize}
\item \verb|Wyckoff_positions|
  \begin{myitemize}
    \description indicates the Wyckoff, letter, multiplicity, site symmetry, position ($3 \times 1$ array), and atom name.
    \type \verb|array of objects|
    \similarto
    \begin{myitemize}
    \item \SPGLIB: \verb|get_symmetry_dataset.wyckoffs| (letters only).
    \item \FINDSYM: in the loop with \verb|_atom| prefix.
    \end{myitemize}
  \end{myitemize}
\item \verb|Bravais_lattice_lattice_type|
  \begin{myitemize}
    \description Bravais lattice of the lattice.
    \type \verb|string|
  \end{myitemize}
\item \verb|Bravais_lattice_lattice_variation_type|
  \begin{myitemize}
    \description lattice variation type of the lattice in the \AFLOW\ standard~\cite{curtarolo:art58}.
    \type \verb|string|
  \end{myitemize}
\item \verb|Bravais_lattice_lattice_system|
  \begin{myitemize}
    \description Bravais lattice system of the lattice.
    \type \verb|string|
  \end{myitemize}
\item \verb|Bravais_superlattice_lattice_type|
  \begin{myitemize}
    \description Bravais lattice of the superlattice.
    \type \verb|string|
  \end{myitemize}
\item \verb|Bravais_superlattice_lattice_variation_type|
  \begin{myitemize}
    \description lattice variation type of the superlattice in the \AFLOW\ standard~\cite{curtarolo:art58}.
    \type \verb|string|
  \end{myitemize}
\item \verb|Bravais_superlattice_lattice_system|
  \begin{myitemize}
    \description Bravais lattice system of the superlattice.
    \type \verb|string|
  \end{myitemize}
\item \verb|Pearson_symbol_superlattice|
  \begin{myitemize}
    \description Pearson symbol of the superlattice.
    \type \verb|string|
  \end{myitemize}
\item \verb|reciprocal_lattice_vectors|
  \begin{myitemize}
    \description reciprocal lattice vectors.
    \type $3 \times 3$ \verb|array|
  \end{myitemize}
\item \verb|reciprocal_lattice_parameters|
  \begin{myitemize}
    \description reciprocal lattice parameters ($a,b,c,\alpha,\beta,\gamma$).
    \type $6 \times 1$ \verb|array|
  \end{myitemize}
\item \verb|reciprocal_volume|
  \begin{myitemize}
    \description reciprocal cell volume.
    \type \verb|float|
  \end{myitemize}
\item \verb|reciprocal_lattice_type|
  \begin{myitemize}
    \description Bravais lattice of the reciprocal lattice (``FCC'', ``BCC'', ``CUB'', ``HEX'', ``RHL'', etc.).
    \type \verb|string|
  \end{myitemize}
\item \verb|reciprocal_lattice_variation_type|
  \begin{myitemize}
    \description lattice variation type of the reciprocal lattice in the \AFLOW\ standard~\cite{curtarolo:art58}.
    \type \verb|string|
  \end{myitemize}
\item \verb|reciprocal_lattice_system|
  \begin{myitemize}
    \description lattice system of the reciprocal lattice.
    \type \verb|string|
  \end{myitemize}
\item \verb|standard_primitive_structure|
  \begin{myitemize}
    \description \AFLOW\ standard primitive crystal structure representation.
    \type \verb|structure object|
  \end{myitemize}
\item \verb|standard_conventional_structure|
  \begin{myitemize}
    \description \AFLOW\ standard conventional crystal structure representation.
    \type \verb|structure object|
  \end{myitemize}
\item \verb|wyccar|
  \begin{myitemize}
    \description \ITC\ conventional crystal structure representation.
    \type \verb|structure object|
    \similarto
    \begin{myitemize}
    \item \SPGLIB: \verb|Spg_standardize_cell(to_primitive=0)|.
    \item \FINDSYM: after \verb|Space Group| line.
    \end{myitemize}
  \end{myitemize}
\end{myitemize}
\hfill\break
The \verb|structure object| lists the following information regarding the crystal structure:
\begin{myitemize}
\item \verb|title|
  \begin{myitemize}
    \description geometry file title.
    \type \verb|string|
  \end{myitemize}
\item \verb|scale|
  \begin{myitemize}
    \description scaling factor of lattice vectors.
    \type \verb|float|
  \end{myitemize}
\item \verb|lattice|
  \begin{myitemize}
    \description row-space representation of lattice vectors ($\mathbf{a},\mathbf{b},\mathbf{c}$).
    \type $3 \times 3$ \verb|array floats|
  \end{myitemize}
\item \verb|species|
  \begin{myitemize}
    \description list of atomic species in crystal.
    \type \verb|array of strings|
  \end{myitemize}
\item \verb|number_each_type|
  \begin{myitemize}
    \description number of atoms for each distinct atomic species.
    \type \verb|array of ints|
  \end{myitemize}
\item \verb|coordinates_type|
  \begin{myitemize}
    \description indicates the coordinate representation (``cartesian'' or ``direct'').
    \type \verb|string|
  \end{myitemize}
\item \verb|atoms|
  \begin{myitemize}
    \description atom information.
    \type \verb|array of atom objects|
  \end{myitemize}
\end{myitemize}
\hfill\break
where the \verb|atom| object contains,
\begin{myitemize}
\item \verb|name|
  \begin{myitemize}
    \description atomic species name.
    \type \verb|string|
  \end{myitemize}
\item \verb|occupancy|
  \begin{myitemize}
    \description site occupancy.
    \type \verb|float|
  \end{myitemize}
\item \verb|position|
  \begin{myitemize}
    \description cartesian or fractional coordinate.
    \type $3 \times 1$ \verb|array|
  \end{myitemize}
\end{myitemize}
\hfill\break
\noindent\textbf{sgdata output.}  
The output from this function is a subset of \verb|edata| containing the space group
and Wyckoff position information.

\newcommand{\Ozolins}{Ozoli\c{n}\v{s}}

\end{document}